\documentclass[acmsmall,screen,nonacm]{acmart}
\usepackage{booktabs} 
\usepackage{subcaption}
\usepackage{comment}
\usepackage{amsmath}
\usepackage{color}
\usepackage{url}
\usepackage{enumitem}
\usepackage{multirow}
\newcommand{\name}{eScope}  
\newcommand{\client}{mobile device}  
\newcommand{\actor}{operator}
\newcommand{\Actor}{Operator}

\newcommand{\Paragraph}[1]{\vskip 3pt\noindent\textbf{#1 }}

\usepackage[normalem]{ulem}
\usepackage{environ}
\ExplSyntaxOn
\seq_new:N \g_appendices_seq
\NewEnviron{Appendix}{\seq_gput_right:No \g_appendices_seq \BODY}
\newcommand\AddAppendices{
  \appendix
  \seq_map_inline:Nn \g_appendices_seq {##1}
}
\ExplSyntaxOff
\allowdisplaybreaks

\AtEndDocument{\AddAppendices}
\usepackage[normalem]{ulem}
\usepackage{etoolbox}
\usepackage{environ}

\usepackage[normalem]{ulem}
\usepackage{etoolbox}
\usepackage{environ}

\AtBeginDocument{%
  \providecommand\BibTeX{{%
    \normalfont B\kern-0.5em{\scshape i\kern-0.25em b}\kern-0.8em\TeX}}}

\acmPrice{15.00}
\acmISBN{978-1-4503-XXXX-X/18/06}

\renewcommand\footnotetextcopyrightpermission[1]{}
\begin{document}


\title[eScope]{eScope: A Fine-Grained Power Prediction Mechanism for Mobile Applications}

\author{Dipayan Mukherjee}
\authornote{Both authors contributed equally to this research.}
\email{dipayan2@illinois.edu}
\author{Atul Sandur}
\authornotemark[1]
\email{sandur2@illinois.edu}
\affiliation{%
  \institution{University of Illinois, Urbana-Champaign}
  \country{USA}
  \postcode{61820}
}

\author{Kirill Mechitov}
\affiliation{%
  \institution{Embedor Technologies}
  \country{USA}}
\email{kmechitov2@embedortech.com}

\author{Pratik Lahiri}
\affiliation{%
  \institution{University of Illinois, Urbana-Champaign}
  \country{USA}
}
\email{plahiri2@illinois.edu}

\author{Gul Agha}
\affiliation{%
 \institution{University of Illinois, Urbana-Champaign}
 \country{USA}}
 \email{agha@illinois.edu}
\renewcommand{\shortauthors}{Sandur and Mukherjee, et al.}


\begin{abstract}
\label{sec:abstract}
    \normalsize Managing the limited energy on mobile platforms executing long-running, resource intensive streaming applications requires adapting an application's operators in response to their power consumption.  For example, the frame refresh rate may be reduced if the rendering operation is consuming too much power. Currently, predicting an application's power consumption requires (1) building a device-specific power model for each hardware component, and (2) analyzing the application's code. This approach can be complicated and error-prone given the complexity of an application's logic and the hardware platforms with heterogeneous components that it may execute on.  We propose \name{}, an alternative method to directly estimate power consumption by each operator in an application. Specifically, \name{} correlates an application's execution traces with its device-level energy draw.  We implement \name{} as a tool for Android platforms and evaluate it using workloads on several synthetic applications as well as two video stream analytics applications.  Our evaluation suggests that \name{} predicts an application's power use with 97\% or better accuracy while incurring a compute time overhead of less than 3\%. 
\end{abstract}




\maketitle

\vspace{-1em}
\section{Introduction}
\label{sec:introduction}

Mobile devices running on batteries have a restricted amount of energy.  Long running, resource intensive applications consume energy at a higher rate (i.e., have a larger power draw), draining the battery faster.  Short battery life is a significant source of dissatisfaction for users of mobile devices ~\cite{BatteryDissatisfaction,SmartwatchUserBattery}.
Since long running applications such as video conferencing and social media apps (e.g., Snap) run for an indeterminate period of time, we do not know what their total energy consumption will be.  However, if we can reduce their power draw, we can extend battery life.

Applications can be thought of as composed of a group of operators which are called in some sequence that depends on the input given to the application.  Some of these operators draw large amounts of power while others draw negligible power.  For many operators, methods exist to reduce the power draw of resource intense operators--thus extending the battery life. For example, reducing bit rate for streaming applications ~\cite{FineABR}, using a simpler and less accurate machine learning model~\cite{s20041176}, controlling configuration parameters such as sampling rate or frame rate ~\cite{potluck,approxtuner}, or off-loading computation to remote servers ~\cite{cuervo2010maui,foggycache,nn-offloading,self}. However, observe that these methods result in a trade-off: reducing energy consumed also reduces the quality of the results.  If we can predict the power draw of long-running applications, we can decide how to manage the trade-off involved. 

Our goal is to predict power at different stages of long running applications.  Observe that the power draw will depend on the input configuration and the characteristics of the platform on which it is executed.

Current methods ~\cite{petra,RECON,cuervo2010maui,hyperpower} for predicting operator-level power consumption require profiling the power consumption of each hardware component, instrumenting the operator source code, and mapping the operator-level software parameters. Hardware-based power models are generated by exercising the hardware components in different operating states, such as utilization and frequency, and measuring the power using external power meters. Fine-grained power attribution requires instrumenting the operator-level source code to estimate each application operator's usage of each hardware component. The power draw is estimated by combining the power profile of each hardware component and their usage by each application operator.

However, existing approaches face several challenges. Creating a hardware-based power model requires manually identifying the parameters impacting the power draw of each component, which becomes challenging with an increasingly diverse set of heterogeneous components (such as multi-core CPU, GPU, NPU, etc.) in different hardware platforms ~\cite{multicore-cpu-power-modeling,petra,eyeriss,ZephyrUsingSoC}. Fine-grained models suffer from additional challenges associated with instrumenting the source code. Lack of source code for proprietary applications can limit the scope of methods utilizing code instrumentation. Also, code instrumentation increases the complexity of the models and creates a maintenance burden for developers ~\cite{ACCBook}. An operator's power consumption depends on the software parameters as well as the content within input video frames (e.g., whether the incoming frame to a face detector contains a face or not); hence only code instrumentation would not provide sufficient information for power estimation. Additionally, an operator's power draw can have a complex and non-linear relationship with the input features (see\S~\ref{sec:escope-challenges}). Hence, identifying all the hardware and software parameters of interest and their dependency across multiple operators required by current methods makes them cumbersome and intractable with the increased diversity and complexity of applications and their underlying hardware components.

We propose  \name{}, an adaptive, resource model-agnostic power prediction method that addresses the challenges of fine-grained power prediction. Our method does not require any manual work for the application profiling or any profiling of the underlying hardware. \name{} also accounts for the impact of input data on each operator's power draw.
Achieving these objectives is challenging because there are
no tools to directly measure the power cost of each operator, and we
do not instrument an operator's code or profile their hardware usage.  Instead, we collect easily accessible device-level power draw information on a mobile device (i.e, battery state-of-charge or SoC) and concurrently monitor the changing composition of the executing operators from different applications.  We use this information to attribute the power drawn in an interval to the active operators. A prediction model, directly mapping operators' execution times to their corresponding power draw, is trained using this data and deployed on the mobile device to provide run-time predictions for power optimization decisions.

We implemented \name{} as a tool for Android platforms in two different languages. We experimentally evaluated \name{} on several benchmarks and two video-streaming applications
across different hardware platforms. Since we cannot individually measure the power draw of each executing \actor{}, we use the \emph{battery drain interval}, i.e., amount of time during which a fixed amount of energy (e.g., 1\% drop in battery SoC on a typical smartphone) is discharged, to measure the accuracy of our model. Our evaluation shows that \name{} can provide high-confidence
predictions with 97\% accuracy of the battery drain interval with multiple executing operators,
with a compute overhead of less than 3\% on the mobile device. It provides prediction accuracy
gains of up to 40\% compared to previous approaches for complex operator execution profiles.
This paper makes the following contributions:

\begin{itemize}[noitemsep,nolistsep,leftmargin=0.15in]
\item A power prediction method for mobile video analytics
  applications which is \textit{transparent} to the application code,
  infers \textit{dynamic} information (thus handling changing input
  streams) and is \textit{device power-model agnostic}.
\item A quantitative analysis of several state-of-the-art machine
  learning models to identify the ones that are best suited to make
  on-device power predictions for applications with complex power
  characteristics such as operators executing in different power
  states depending on their input, while incurring negligible
  overhead.
\item A fully-functional tool called \name{} supporting mobile
  applications across two different programming languages: (1)
  SALSA~\cite{SalsaOnAndroid} which supports actor-based programming
  and provides native support for code migration (to be leveraged for
  future extensions of \name{}), (2) Kotlin~\cite{kotlin-actors} a
  language that is popular for building mobile code.  We evaluate our
  tool using several benchmarks to show that it accurately predicts
  operator power costs and incurs low overhead on the mobile device.
\end{itemize}

\section{Related Work}
\label{sec:related-work}

Given the importance of power management in mobile phones, there has been significant interest in modeling the power consumption of various applications and predicting the phone's battery life. In this section, we categorize the related work in terms of techniques used for measuring, modeling, and attributing the power consumption of the mobile phone.

\noindent \textbf{Power measurement methods}: Mobile device power can be easily measured using external power meters \cite{demo:externalpowermonitor, monsoonpowermonitor} as demonstrated in recent works \cite{PowerScope, JouleWatcher, eProfSystemCall, RECON}, or using specialized hardware like OCV-based fuel gauge chips used in Sesame ~\cite{Sesame}. Using power meters or specialized hardware for modeling applications' power consumption is limited to laboratory settings and impractical for broader adaptation.

Automated measurement solutions use the remaining battery capacity on the device SoC or battery voltage readings (e.g., V-edge \cite{Vedge}) to measure power consumption. V-edge models power consumption by mapping the instant battery dynamics with battery levels. This approach requires frequent calibration based on the age and wear \& tear of the battery. We use battery SoC to measure the power consumption as used by previous approaches \cite{ZephyrUsingSoC, BatteryTraces}, allowing us to model outside the laboratory setting and avoid the limitations of other mentioned approaches. 

\noindent \textbf{Power Modeling}: Resource-based power modeling involves leveraging the utilization of various hardware components (e.g., CPU, network) to generate models which map them to power costs. PowerTutor ~\cite{PowerTutor} models power by mapping different hardware power states to SoC, while DevScope ~\cite{DevScope}, AppScope ~\cite{AppScope}, and MARVEL ~\cite{MARVEL} isolate shared hardware resource usage among resources. These resource-based utilization models provide a coarse-grained model. However, coarse-grained application-level models cannot capture the impact of software parameters(such as frame rate and input resolution) and input content on the power draw.

A hybrid approach has been adopted by recent works~\cite{RECON,neuralpower,eProfSystemCall} which use application-based data in addition to the resource-based power model to address the limitations of resource-based power models. RECON ~\cite{RECON} uses application component data in addition to the resource utilization model, while eProf ~\cite{eProfSystemCall} tracks system calls to model tail energy. Hybrid models provide fine-grained, operator-level power consumption information. These models require hardware parameters in addition to software parameters to capture the dynamic power state of an operator based on input data.  For example, the power draw of the facial recognition operator will increase depending on the existence of a face in the input frame. However, with the availability of a variety of hardware components (such as multi-core CPU, GPU, memory, network, NPU, etc.), finding the relevant feature for each hardware takes time and effort. Instead, our proposed method \name, utilizes alternative features such as operator execution time profiled with high accuracy, to distinguish between multiple power states resulting from the operator processing different types of input data.

Application-level software parameters, such as frame rate, image resolution, and precision of weights for a deep learning model within an operator, have been used to build operator-level power models~\cite{RECON,neuralpower}. These methods require developers to identify model parameters manually, which makes exploring large and exponential parameter space of parameters in large video analytics code-bases challenging ~\cite{chameleon-video-analytics,llama}. These models depend on software parameters such as frame rate, resolution, etc. Our work addresses the impact of input video frame content on power consumption, without relying on the large space of software parameters.

\noindent \textbf{Power Attribution}: Modern SDK tools (such as BatteryStats and BatteryHistorian ~\cite{battCheck} for Android and, Instruments~\cite{iosIntrument} for iOS ) use resource-based models for estimating energy consumption during execution. These tools provide coarse-grained information unsuitable for fine-grained operator-level power models. Recent works ~\cite{neuralpower, hyperpower} have proposed developing fine-grained power models using resource-specific models for newer resources (such as GPU and TPU) for specific workloads, such as CNN or neural networks. However, these methods require re-modeling for each iteration of new hardware. Our approach overcomes the limitation of re-modeling by adopting a resource-agnostic model.

Current fine-grained power prediction relies on building hardware resource models and program instrumentation~\cite{cuervo2010maui}. Instrumentation is challenging as increasing code complexity burdens developers and increases maintenance costs. Static analysis allows for automated instrumentation \cite{automated-code-instrumentation}. However, these methods still require dynamic analysis to quantify loop operations, which relies on resource-specific models ~\cite{polarized-session-types,resource-aware-session-types}. We provide a detailed analysis of the limitations of instrumentation and hardware models in \S~\ref{sec:problem-def}. \name\ proposes an approach that does not suffer from these limitations, as we do not use code annotation or any resource-specific modeling.

\section{Motivating scenarios}
\begin{figure}[!b]
    \centering
    \includegraphics[width=0.85\textwidth]{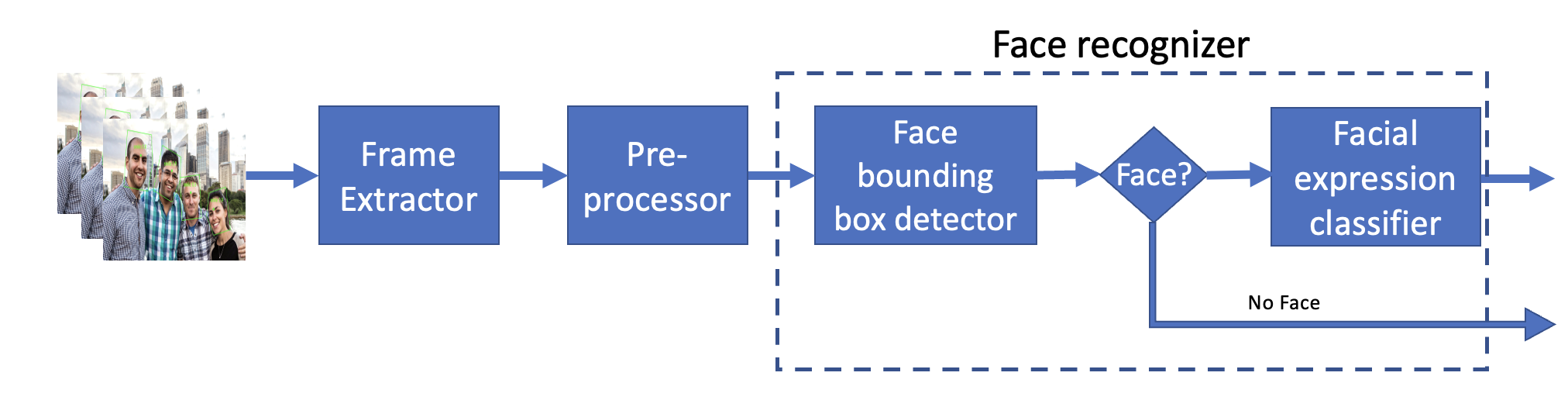}
    \caption{\normalfont An application can be divided into operators that are called in some sequence depending on the input configuration.  \Actor{}s in a face recognition application.}
    \label{fig:face-detection-pipeline}
\end{figure}

Mobile video analytics applications, such as video conferencing apps, augmented reality games (e.g., Pokemon Go), and face filters on social media (e.g., Snap), are some of the widely used long running, resource-intensive applications. These applications comprise compute-intensive operators such as  face detection, facial expression classifier, as shown in Fig \ref{fig:face-detection-pipeline}. An operator's power consumption depends on various factors, such as the software parameters of the application (frame rate and resolution of input), the content within input video frames (e.g., whether the incoming frame to a face detector contains a face or not), the executing hardware platform, and the current state of the battery.

 To accurately predict the power consumption of these applications we need to understand the fine-grained (operator-level) power consumption of the executing application accurately. Existing techniques for \actor{}-level power predictions are challenging to use with the increased diversity and complexity of applications and their underlying hardware components.

\subsection{Challenges}
\label{sec:escope-challenges}
\begin{figure}[htp]
    \centering
    \includegraphics[width=0.5\textwidth]{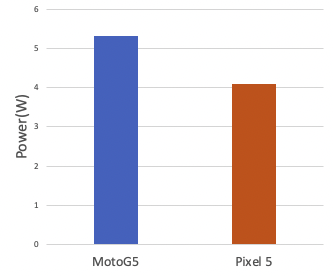}
    \caption{\normalfont Variation in power consumption of a fixed workload on two different phones}
\label{fig:hw-variation}
\end{figure}

Existing power prediction techniques rely on modeling \actor{}-level power draw based on their compute resource utilization. With hardware advancements, developing analytical power models is becoming increasingly challenging. Simple utilization-based linear models do not apply various resources, such as multi-core CPUs that do not capture the effect of CPU idle power ~\cite{multicore-cpu-power-modeling}, GPU where power depends on additional factors like memory bandwidth, frequency of shader cores~\cite{GPUPowerModel,gpupowermodelsurvey}. Previous works have highlighted the challenges of building complex, analytical power models for each new hardware component~\cite{ZephyrUsingSoC,petra,energy-mobile-video-streaming,neuralpower,multicore-cpu-power-modeling}. Moreover, the same application shows high variation in its power costs when run on different mobile devices. To show this hardware dependency, we run ML Kit~\cite{mlkit} based face detector app on two different phones: \textbf{Pixel5}(Chip: Snapdragon 765G) and \textbf{MotoG5} (Chip: Snapdragon 430).  Despite the same input, two phones exhibit high variation in power draw and energy consumption of the application (4.1 W
and 5.3W
respectively) as shown in Fig~\ref{fig:hw-variation}. Each \actor{}'s power draw is sensitive to the hardware device on which it runs and the nature of input video content it is processing.

\begin{figure}
    \centering
    \includegraphics[width=0.65\textwidth]{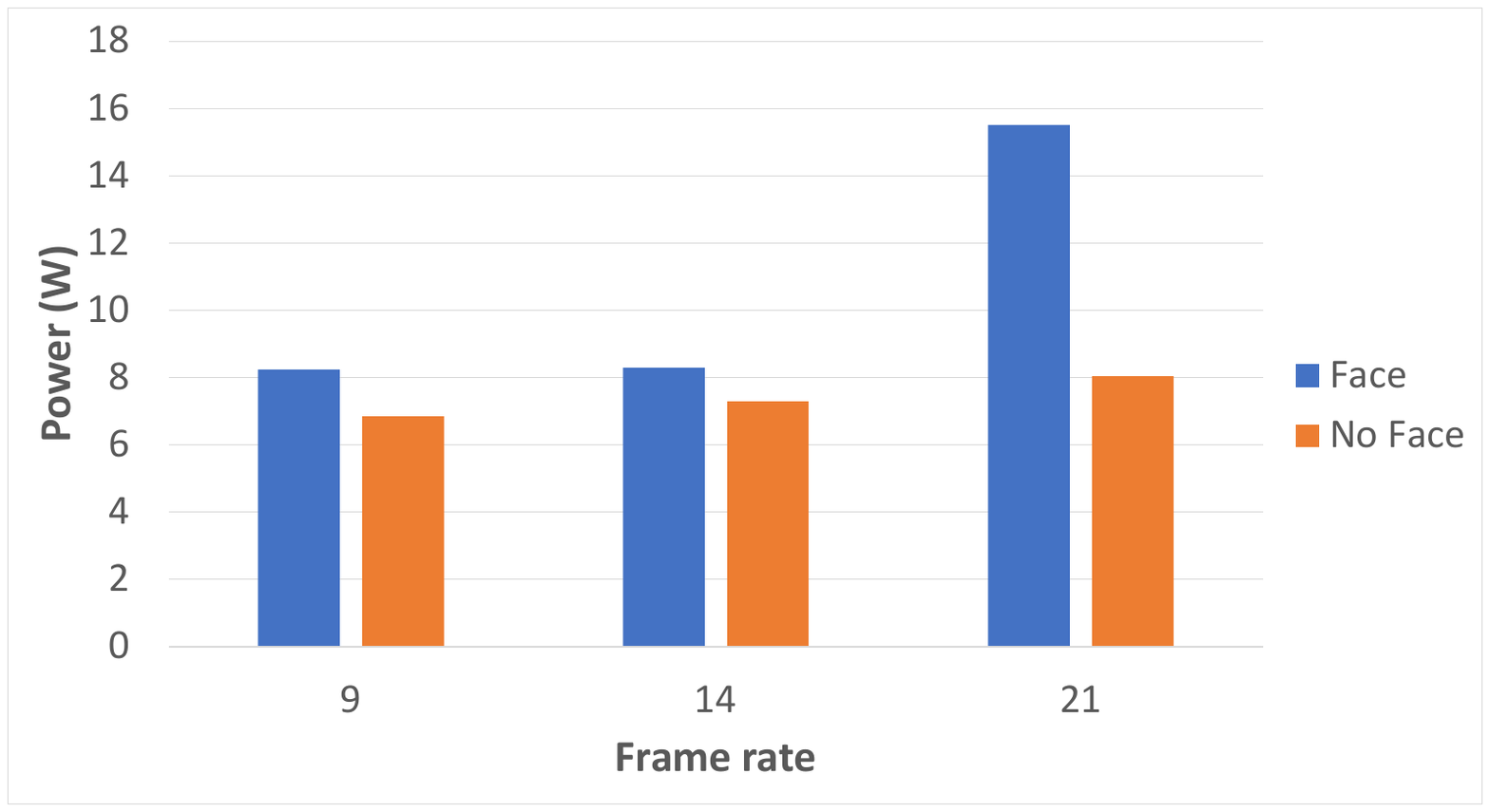}
    \caption{\normalfont Input sensitivity to the frame rate on the power draw of a video analytics application }
\label{fig:input-dependency-face-recognizer}
\end{figure}

\Paragraph{Input-dependent power consumption.} To quantify the dependency of power on input frame content, we vary the input to the face-recognition application (Fig. ~\ref{fig:face-detection-pipeline}) running on \textit{Pixel 5}. The application consists of two \actor{}s: \textit{face bounding box detector} and \textit{facial expression classifier}. If a face is detected in the frame by first \actor{}, it is sent to the classifier; otherwise discarded. The power cost for the two input video streams, with and without faces,  shows a significant variance (13\% to 92\%) over different frame sampling rates (Fig~\ref{fig:input-dependency-face-recognizer}). This variation can be attributed to the conditional logic of the application, where frames with faces additionally run the resource-intensive classifier. We also observe that increasing frame rate increases the power cost, indicating that it is an essential factor impacting power.

Finer-grained \actor{} definitions (i.e., different \actor{}s for the face detector and classifier) could remove input dependency by mapping each \actor{} to a single power cost. However, \name{} has no control over the granularity of \actor{}s. Coarse-grained \actor{}s result in sequential \actor{} execution per frame, while fine-grained \actor{}s can cause significant control/scheduling overheads. Thus,\actor{}s are designed based on business logic and performance considerations. Moreover, third-party library developers commonly define conditional logic within  \actor{}s for end-user applications (e.g., avatars for users in video chat). Our observations on input dependency of \actor{} power draw have been corroborated by recent works~\cite{llama}. We need a power prediction mechanism that considers the input dependency of \actor{} costs.

\section{System Model \& Problem Definition}
\label{sec:problem-def}

This section describes our system models and assumptions and defines the power prediction problem.

\Paragraph{Application Model: } Applications are represented as directed acyclic graphs (DAG). The DAG vertices denote stream operations mapped to a specific application function (e.g., face detector), and the edges denote data flow dependencies between \actor{}s. Scheduling is done at the \actor{}-level, enabling us to trace \actor{} execution from the underlying scheduler, without code-level changes.

\Paragraph{Device Model:}
\label{sec:energy-measurement-proposed}

We consider battery-driven mobile devices which typically expose a battery status API for estimating the state-of-charge (SoC), which is a measure of the remaining battery charge as a percentage of the total. SoC helps measure the device-level energy consumed by executing applications. The device OS exposes an API for querying the SoC and uses a hardware component called \textit{battery fuel gauge}~\cite{survey-energy-profiling}. The fuel gauge has an estimation error within 2\%~\cite{maximdatasheet}. 

A client can register with the battery status API, which notifies its listeners whenever there is a change in the SoC estimate. The obtained estimate is converted to energy consumed by the following equation~\citep{PowerTutor}:
\begin{equation}
P * (t_1 - t_2) = E * (SoC(V_1) - SoC(V_2))
\label{eq:PowerSocEq}
\end{equation}
 \hspace{0.2 in} where $P$ is the average power consumption in time interval $[t_1,t_2]$, $E$ is the rated battery energy capacity, and $SoC(V_i)$ is the battery state-of-charge at voltage $V_i$ ($i$ is 1 or 2). The battery's duration at an SoC level is inversely proportional to the average device power consumed. The lowest granularity for SoC change is 1\% for most mobile devices.


We use the SoC estimates as it provides an easily accessible (without a power meter), cross-platform (uses the phone's built-in battery sensor) solution for measuring power on the mobile device~\cite{ZephyrUsingSoC}. However, these estimates can be extremely noisy, and battery characteristics change due to temperature and age~\cite{AppScope, LeeModelGen,PowerTutor}. In \S ~\ref{sec:energy-estimation-model}, we discuss pre-processing steps used to extract the relative difference in \actor{} power estimates for offloading purpose.  In \S~\ref{sec:energy-estimation-model}, we discuss our prediction models.





\Paragraph{Power Consumption Model:}. \Actor{}s may perform computation and/or communication tasks depending on the programmed application logic. Such tasks contribute to the power drawn by the scheduled \actor{} which executes them. SoC measurements to estimate the power drawn in a given battery drain interval, correlate with the scheduled \actor{}s within that interval. To validate this, we ran an experiment with an \actor{} uploading different sizes of video data to a remote server over time, and recorded SoC changes during that time. Based on network interface power models~\cite{PowerTutor, cuervo2010maui, IMCM-full}, we expect the power consumption to increase with uploaded data size. 
An \actor{} performing asynchronous send/receive operations causes energy drain which can be measured by changes in SoC during that time interval. Power consumption models based on usage of other resources such as CPU~\cite{PowerTutor, cuervo2010maui}, memory~\cite{eyeriss-memory-neuralnets,survey-energy-profiling} also indicate that the \actor{}'s resource usage corresponds to its execution power cost. Resource usage results in load which influences the observed battery output voltage and is reflected in the SoC measurements.  

\Actor{}s can also exhibit complex power usage characteristics, residing in multiple power states depending on the video input. E.g., an \actor{} in a face recognition application could pre-process video frames (which incurs CPU/memory cost) and, based on pre-processing output, either upload the frame for matching against a face database (incurs network cost on mobile device) or discard it (no energy expended)~\cite{IMCM-full, RECON}. Figure~\ref{fig:input-dependency-face-recognizer} also shows that input parameters such as frame rate can affect the \actor{} power cost. Our power consumption model considers that same operator can exist in different power states depending on the input data.   

\Paragraph{Problem Definition: } 
Consider a set of one or more application \actor{}s scheduled on the \client{} within a \textit{battery discharge interval} (or \textit{battery drain interval}), which is the amount of time during which a fixed amount of energy (e.g., 1\% drop in battery SoC on a typical smartphone) is discharged due to executing applications on the \client{}. When an \actor{} is scheduled to run, it is considered to be \textit{active} until it completes executing in the current scheduling interval. An \textit{execution segment} is a time duration during which the composition of active \actor{}s does not change. Thus, execution segments are non-overlapping. 

A sequence of segments is represented as $SS_{i} = \{s_{j}^{i}\ \vert\ j \ \in\ \{1,...N_{S}^{i}\},\ s_{j}^{i} \subseteq V_{o}\}$  where $s_{j}^{i}$ consists of the set of active \actor{}s in the $j^{th}$ segment of $i^{th}$ discharge interval, $N_{S}^{i}$ is the total number of segments in the $i^{th}$ discharge interval and $V_o$ is the set of application \actor{}s. $\mathcal{E}_{j}^{i}$ is the actual (i.e., ground-truth) energy consumed by segment $s_{j}^{i}$ during time $t_{j}^{i}$ that it was executing. We define an energy map from segments to their energy costs as $EMap : SS_{i} \times \mathbb R_{\ge 0} \to \mathbb R_{\ge 0},\ EMap(s_j^{i},t_j^{i}) = ecost_j^{i}$, where $ecost_j^{i}$ is the energy consumption of segment $s_j^{i}$ executing for duration $t_j^{i}$. We assume each discharge interval corresponds to 1\% battery SoC drop, so $i \in [1,100]$. The energy prediction problem is to find $EMap$ for which the mean-squared error of actual and estimated energy costs of segments in each discharge interval $i$ is minimized:
\begin{equation}
\textrm{Mean-squared error} = \frac{1}{N_{S}^{i}}\sum_{j=1}^{N_{S}^{i}} (\mathcal{E}_j^{i} -  EMap(s_j^{i},t_j^{i}))^2 
\label{eq:objective-escope}
\end{equation}
Our goal is to use coarse-grained device-level energy values and attribute them to fine-grained application \actor{}s. 
\Paragraph{Challenges: }
Existing works ~\cite{PowerTutor,cuervo2010maui,IMCM-full} profile the hardware resources with an external power meter~\cite{survey-energy-profiling} to build a resource utilization-based power model~\cite{PowerTutor}. \Actor{}s are instrumented for their resource use, and the power model is used to determine their energy costs. However, such power monitoring techniques have several disadvantages. They rely on external power meters which are not practical to carry with the \client{}. \textit{Source code instrumentation} burdens application developers, increases code complexity and maintenance costs~\cite{ACCBook}. Resource power models exhibit \textit{high prediction error}~\cite{HardwareModelsDoNotWork}, and need to be rebuilt for new hardware components~\cite{GPUPowerModel,tpu-power-model,multicore-cpu-power-modeling}. Power costs of \actor{}s depend on input content and other software parameters such as frame rate and resolution. Recent works~\cite{RECON,hyperpower} have incorporated software parameters into their power models, but they are generally simple linear models or tightly integrated to specific platforms (e.g., NVIDIA GPU toolchain). These models cannot capture the effect of the same \actor{} executing in multiple power states depending on input frame content.

To overcome the above disadvantages, we propose the following goals for our solution: 
(i) \textit{Empirical and resource model-agnostic} solution which does not require manual work to build analytical power models for each of the hardware or application components., (ii) \textit{Easy to adopt / Zero-effort}, not requiring any changes to application code, (iii) \textit{Dynamic}, to support complex runtime resource usage behaviors depending on input frame content, (iv) \textit{Lightweight}, so that resource overhead on the \client{} is minimal.

A significant challenge for the power prediction problem is that \client{}s do not provide a mechanism for direct power measurement of each \actor{}. Battery SoC is easily accessible for querying on the device but is extremely noisy and coarse-grained. Moreover, since we do not have access to application code, behavior of each \actor{} and its energy and power use is hidden, making it challenging to attribute coarse-grained battery SoC to the executing \actor{}s. 

\section{Proposed Approach}
\begin{figure}[htp] 
  \centering
    \includegraphics[width=0.9\textwidth]{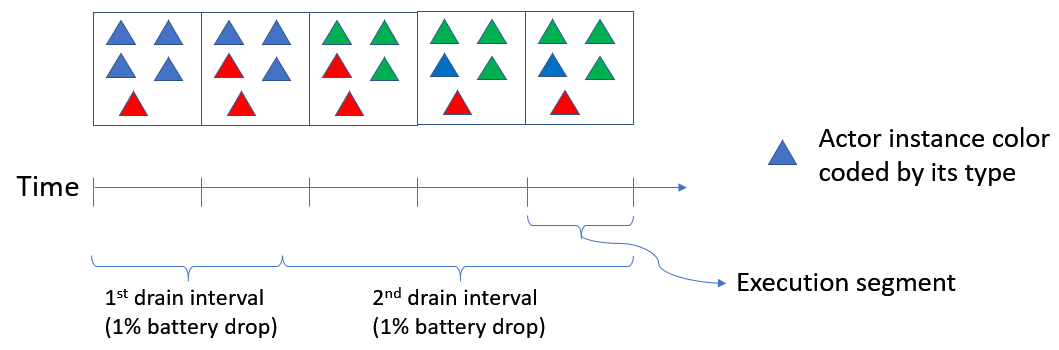}
    \caption{\normalfont A snapshot of execution segments and discharge intervals for \actor{}s executing on the mobile device.}
    \label{fig:execution-model}
\end{figure}

Our approach utilizes the battery status API exposed by the operating system to estimate the remaining battery charge (SoC) as a percentage of the total. We instrument the system run-time scheduler to monitor the time duration of active application \actor{}s on the device. We observe that the composition of active \actor{}s varies with each battery SoC level, depending on the input to the streaming application (e.g., downstream \actor{}s after a filter will not execute if the frame is filtered out). Additionally, the time duration of \actor{}s executing in a discharge interval can change. We use these insights to train prediction models for the power cost of active \actor{}s in each discharge interval.
\subsection{Execution segment}
\label{sec:execsegment}

Execution segments are obtained by tracking the scheduling and completion of the \actor{}s in the application. We instrument the underlying runtime scheduler to store \actor{} details such as type(e.g., filter, feature extractor) and the number of instances of each type in a map, while avoiding changes to the application. The runtime exposes an API to be polled for querying segment information. The time interval between consecutive polls is configurable and is referred to as \textit{polling interval}. Too large an interval can increase the memory size of segment data stored by the runtime while too small an interval increases the polling overhead. We find that a 1 second polling interval works well in our experiments. 
\subsection{Power attribution}
\label{sec:power-attribution}
\name\ power predictor maps \actor{}s directly to their power costs (no analytical model required). In Eqn~\ref{eq:objective-escope}, we defined our minimization objective for estimating the power costs of \actor{}s within each discharge interval. Each discharge interval is broken down into a sequence of execution segments, thus providing one training data point. Fig~\ref{fig:execution-model} shows how the execution segments are aligned with battery discharge intervals. 
 
The sequence of execution segments in the $i^{th}$ discharge interval $SS_i$ consists of either active \actor{}s executing (referred to as \emph{active segments}) or no active \actor{}s during which time the device is idle (referred to as \emph{idle segments}). The occurrence of active and idle segments depends on the nature of the incoming video stream and scheduling behavior of the underlying runtime. As shown in Figure~\ref{fig:execution-model}, an active segment may consist of one or more active \actor{}s executing concurrently. We represent the $j^{th}$ segment $s_{j}^{i} \in SS_i$ as $s_{j}^{i} = \{o_k\ |\ k \in [1,M], o_{k}\ \textrm{is an active \actor{} in segment}\ s_{j}^{i}\}$. If $s_{j}^{i}=\{\}$ then it is an idle segment. Based on Equation~\ref{eq:PowerSocEq}, if we fix the SoC at 1\%, then the length of a discharge interval is inversely proportional to the expected power consumed during that time interval. We use $E_{1\%}$ to denote the fixed energy consumed in a discharge interval. The energy consumption within the $i^{th}$ discharge interval is as follows: 
\begin{align*}
    E_{1\%} &= \sum_{j=1}^{N_{S}^{i}} EMap(s_{j}^{i}, t_{j}^{i})\\  
     &= \underbrace{\sum_{j \in [1, N_{S}^{i}] : s_{j}^{i} \neq \{\}} EMap(s_{j}^{i}, t_{j}^{i})}_\text{Active Energy} + \underbrace{\sum_{j \in [1, N_{S}^{i}] : s_{j}^{i} = \{\}} EMap(s_{j}^{i}, t_{j}^{i})}_\text{Idle Energy}\\ 
    &= \sum_{j \in [1, N_{S}^{i}] : s_{j}^{i} \neq \{\}} p(s_{j}^{i}) \cdot t_{j}^{i} + p_{idle} \cdot (T_{1\%}^{i} - \sum_{j \in [1, N_{S}^{i}] : s_{j}^{i} \neq \{\}} t_{j}^{i}) \stepcounter{equation}\tag{\theequation}\label{eq:energy-segmentwise-withpower}
\end{align*}
\noindent where $p(s_{j}^{i})$ is the power consumption of executing the $j^{th}$ segment of $i^{th}$ discharge interval, $p_{idle}$ is the constant idle power on the mobile device, $T_{1\%}^{i}$ is the duration of $i^{th}$ discharge interval. Re-arranging terms in Equation~\ref{eq:energy-segmentwise-withpower}, gives us the following equation: 
\begin{equation}\label{eq:final-energy-eq-escope}
    \begin{aligned}
        T_{1\%}^{i} 
        &= \frac{E_{1\%}}{p_{idle}} + \frac{\sum_{j \in [1, N_{S}^{i}] : s_{j}^{i} \neq \{\}} (p_{idle} - p(s_{j}^{i})) \cdot t_{j}^{i}}{p_{idle}}
    \end{aligned}
\end{equation}

According to Eq \ref{eq:final-energy-eq-escope}, if $p(s_{j}^{i})) > p_{idle}$ and if more the segments in discharge interval that we observe \actor{}s within segment $s_{j}^{i}$, the smaller the total number of segments in that interval. Given the traces for battery SoC and application traces, we can obtain the value of $T_{1\%}^{i}$ for the $i^{th}$ discharge interval, active segments along with their durations in segment sequence $SS_{i}$ and the \actor{}s in each segment. This enables us to estimate parameters of Equation~\ref{eq:final-energy-eq-escope}. Note that for placement purposes, it suffices to estimate these parameters, which compute for a given segment $s_{j}^{i}$, the coefficient $\frac{p_{idle} - p(s_{j}^{i})}{p_{idle}}$. We refer to the absolute value of this coefficient as the \textit{relative power} for the corresponding segment.

\subsection{Power prediction model}
\label{sec:energy-estimation-model}

In \S~\ref{sec:escope-challenges}, we observed that power cost is sensitive to the input stream content. An \actor{} can have multiple power states owing to input variability. We propose to use the execution time of each \actor{} within a discharge interval, as an input feature to our prediction model. Execution time captures changes in processor utilization of compute-intensive \actor{}s due to changes in incoming video content and the effect of software parameters such as frame rate, the use of a high vs. low accuracy DNN model within a ML operator, frame resolution, etc. We rely on high quality training datasets (i.e., data covering a wide range of execution conditions with varying input content and parameters such as frame rate) to train our prediction models. To handle the variability in \actor{} power consumption across different target hardware, training is done separately for each mobile device. In evaluation, we show that the custom, device-specific models provide high accuracy power predictions for video analytics applications.
\begin{figure}
    \centering
    \includegraphics[width=0.75\textwidth]{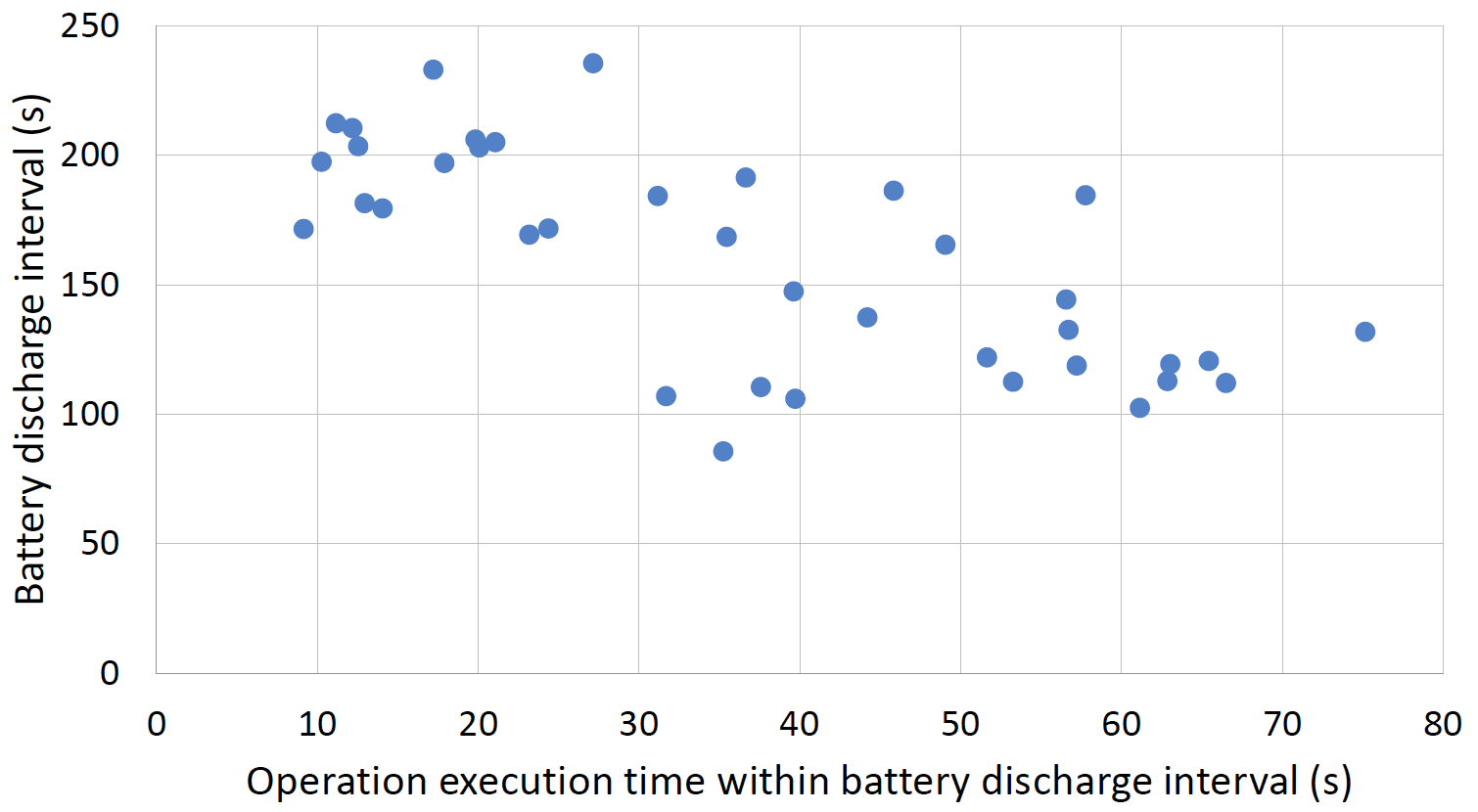}
    \caption{\normalfont Sample training data for \name{} from a video conferencing application, with the \actor{} execution time and battery discharge interval being the independent and dependent variables respectively, from Eq~\ref{eq:final-energy-eq-escope}.}
    \label{fig:escope-video-conference-non-linear-motivate}
\end{figure}

Linear models are popular for predicting \actor{} power costs using hardware or software parameters~\cite{PowerTutor, RECON, hyperpower}. In \S~\ref{sec:power-attribution}, we observed that linear power models work best for \actor{}s processing a fixed input stream, thus running in a single power state. For changing input, we need complex, non-linear functions for mapping \actor{} power costs to their execution time. To demonstrate this, we ran experiments using a mobile video conferencing application (see Figure ~\ref{fig:escope-video-conference-app}) that supports background subtraction and blurring operations during a video call. The input stream is generated at different frame resolutions while  changing its content with only some frames having a face, thus triggering the execution of segmentation within the background subtraction \actor{} only for those frames. We focus on the following target scenario for our application use: \textit{higher sampling rate} when battery SoC is high and \textit{lower sampling rate} for lower battery SoC levels, such that the user quality of experience is maximized, subject on energy constraints. Since the input feature space can be large to cover all use cases, a developer can provide a representative set of training inputs focused on this target scenario for the application. Fig ~\ref{fig:escope-video-conference-non-linear-motivate} plots the training dataset for our video conferencing application, based on the X and Y-variables in our power prediction model described in Eq ~\ref{eq:final-energy-eq-escope}. The battery discharge interval varies widely (100-200 seconds) with a correlation coefficient of -0.7 with the \actor{} execution duration. In \S~\ref{sec:evaluation}, we will study complex prediction models (e.g., support vector regressor ~\cite{svr}) that can perform significantly better than linear regressors. The background subtraction \actor{} can exist in multiple power states as it executes the expensive segmentation step within background subtraction \actor{} only if the input frame contains a face. Thus, \name{} needs to handle complex training data depending on target use cases for the application.

Before looking into different prediction models considered by \name{}, we discuss pre-processing steps to clean up the training data and reduce noise/measurement errors.

\noindent \textbf{Pre-processing}: Power measurements collected from the mobile device are noisy and can affect prediction model accuracy. There are multiple sources of noise: (1) For the same fixed compute load, discharge interval lengths can have high variance across multiple samples. However, under high load, we observe the variance to be lower, making it easier to isolate the more interesting, compute-intensive \actor{} costs, (2) Under high compute load, an \actor{} gets scheduled in a segment but gets CPU cycles only in the next segment or the thread polling the runtime API to query the segments may not be scheduled, increasing the memory overhead for storing segment data and potentially missing \actor{} information.   

We perform the following pre-processing steps to alleviate effects of measurement noise: (1) We compute the expected number of polls in a discharge interval and discard intervals in which the difference between actual and expected number of polls within the interval is greater than a threshold (e.g., 10\%), (2) To handle high variance discharge intervals, we bucketize segment durations and map each discharge interval to the bucket duration to which it corresponds. Discharge intervals which are outside two standard deviations for a bucket duration, are filtered out. Finally, before reporting the accuracy of a trained model, we examine residual plots and Cook's distance~\cite{cooks-distance} to discard outlier samples.

\noindent \textbf{Training and Prediction}: After pre-processing, we have segment-level active actor set data for each discharge interval. This data is uploaded from the mobile device to the remote \name{} server for training a  bag of prediction models. \name{} continues to collect active actors and discharge intervals in the background, improving the power model and redeploying it when model parameters change due to dynamic resource conditions or application updates.

Eqn~\ref{eq:final-energy-eq-escope} suggests that a linear regressor is sufficient to model the power consumption of active \actor{}s. Several real-world applications, including web browsers~\cite{RECON} and real-time face detectors (\S~\ref{sec:evaluation}) show more complex behaviors. The profiling results of a face detector based on ML Kit library~\cite{mlkit}) show that multiple power states depend on the duration for which the components execute. Training models that capture non-linear dependencies between segment power costs and their execution durations would provide better accuracy for such complex components exhibiting multiple power states. We analyze a comprehensive set of ML models for their accuracy and resource overhead for making power predictions on the mobile device: 
\begin{enumerate}[label=\textbullet,leftmargin=0.3cm]
\item
Linear regression: Based on Eq.~\eqref{eq:final-energy-eq-escope}, linear regressor can model energy behaviors of active actors executing in fixed power states. 

\item Generalized additive models (GAMs)~\citep{GAM} are more flexible than linear models, since they estimate functions of the predictor variables which are connected to the dependent variable via a link function. But they require manually selecting features and choosing optimal smoothing parameters for those features.
\item Support Vector Regression (SVR) ~\citep{svr} is useful when the dimensionality of feature space representation is much larger than number of training observations. It is suited for handling limitations of linear functions in high dimensional feature spaces.  
\item Tree-based regressors are considered black box techniques and hard to interpret. Random forest (RFR)~\cite{RFR} and gradient boosting regressors (GBRs)~\cite{gbr} are ensemble learning methods which are easy to train and capture complexities in the data very well. Unlike RFR, GBR is a boosting technique which can outperform RFR if tuned well but has tendency to overfit if the data is noisy.
\item Neural Network (NN) are extremely flexible and can learn important features from complex input data. Even though they are difficult to train and require larger amounts of data, they can be computationally less expensive than techniques such as RFR.    
\end{enumerate}

We use nested cross-validation (Nested CV)~\cite{NestedCV} to evaluate each of our prediction models. The best prediction model after training is deployed back to the mobile device for making on-device and runtime power predictions. So \name\ identifies the best model in terms of prediction accuracy which also minimizes overhead on the mobile device. 

\section{Implementation}\label{sec:monitoring}
\begin{figure}[h]
  \centering
  \includegraphics[width=0.48\textwidth]{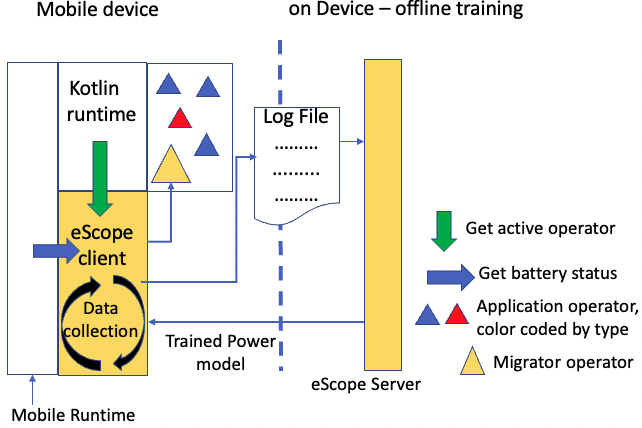}
    \caption{\normalfont System architecture 
    }
    \label{fig:architecture}
\end{figure}

\name\ consists of two components (Fig. ~\ref{fig:architecture}) : (1) a client which runs on a mobile device and (2) a server on a remote machine. The client periodically queries GetActiveOperatorsAPI implemented in the language runtime scheduler, which returns the active \actor{}s within the segment from the last polling interval. Furthermore, the GetBatteryStatus API provides the battery SoC information. Power prediction models are trained on the server using the logs collected at the device during the training phase. The best-identified model on the client data is used application \actor{}-level prediction.


\begin{figure}
\begin{minipage}[b]{0.40\textwidth}
 \includegraphics[width=\textwidth]{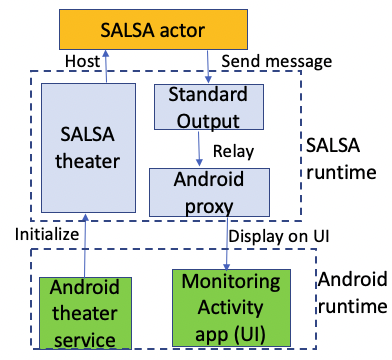}\\
\subcaption{\normalfont SALSA runtime for \name{}.}
    \label{fig:salsa_android}
\end{minipage}%
\begin{minipage}[b]{0.4\textwidth}
\includegraphics[width=\textwidth]{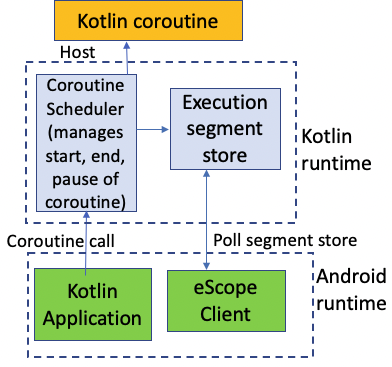}\\
\subcaption{\normalfont Kotlin runtime for \name{}.}
    \label{fig:kotlin_android}
\end{minipage}%
 \caption{\normalfont Changes in SALSA and Kotlin runtime to implement \name{}.}
\end{figure}

We implemented \name{} in two different language runtimes: SALSA~\cite{SalsaOnAndroid} and Kotlin~\cite{kotlin-actors}. We chose those two languages as SALSA allows actor abstraction to implement \actor{}s, and Kotlin is a popular and mainstream programming language with strong support for Android development. We chose Android~\cite{android} platform for our mobile devices due to its easy programmability and support for a wide variety of tools. 

SALSA is easily ported to Android as it compiles to Java byte code. We leveraged prior research~\cite{imais-taskoffloading-android-salsa} to model the architecture (Fig~\ref{fig:salsa_android}) for SALSA actor interaction on Android. The AndroidTheaterService and AndroidProxy service enables the hosting, migration, and communication of SALSA actors with Android drivers.

Kotlin coroutines are suspendable computations that allow code to run concurrently without being tied to a thread, conceptually similar to a lightweight thread. Coroutines are used to implement the video analytics \actor{}s. We modify the underlying language runtime to implement \name{}, allowing us to extract information of coroutine and calling \actor{}. We add the coroutine information, such as runtime and active/inactive, to a dictionary keyed by its unique \textit{coroutineID} that allows disambiguating multiple calls from the same coroutine. \name{} uses the exposed API to query \actor{} information for training and infering power cost.

\begin{figure}[!b]
  \centering
    \includegraphics[width=0.75\textwidth]{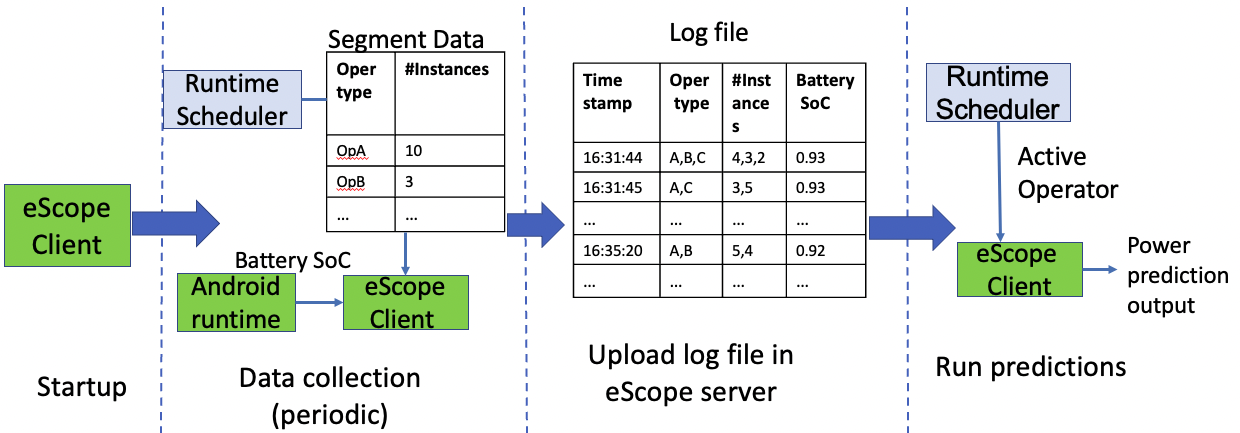}
    \caption{\normalfont Phases of execution for \name{} client,
   }
    \label{fig:monitoring-app}
\end{figure}

The \name\ Android app has multiple phases (Fig ~\ref{fig:monitoring-app}). It launches the language runtime scheduler for \actor{}s to run and periodically polls segment and battery SoC~\cite{AndroidBatteryManager}. The data is added to a log file with a timestamp and uploaded to the server after the training phase. The app then downloads the trained model from the server for on-device power prediction. Scikit-learn~\cite{scikit} python module is used to implement the prediction model, and Keras~\cite{keras}  with Tensorflow~\cite{tensorflow}  were used for neural networks. Nested CV~\cite{NestedCV}  using grid search~\cite{gridsearch} hyperparameter tuning was implemented for comparison. The best model is then exported to a PMML file on the training server for the client to use.

\section{Evaluation}
\label{sec:evaluation}
In this section, we evaluate \name\ for its effectiveness in monitoring power consumption across a wide range of application and execution profiles.

\vspace{-0.7em}
\subsection{Experimental setup}
We ran experiments on the following mobile devices: (1) Samsung Galaxy Nexus (GTI9250)- Android 4.1.2, 2 ARMv7 CPU, 710.8 MB RAM, Li-Ion battery capacity 1750mAh. Its removable battery helped obtain ground truth for the power drawn by the workloads when directly connected to a power supply, (2) Samsung Galaxy S6 (SM-G920T)- Android 7.0, Octa-core (Cortex-A57 \& Cortex-A53) CPU, 3 GB RAM, Li-Ion battery capacity 2550 mAh. (3) Google Pixel 5 - Andriod 12.0, Qualcomm Snapdragon 765G, with Octa-core (1 × 2.4 GHz Kryo 475 Prime \& 1 × 2.2 GHz Kryo 475 Gold \& 6 × 1.8 GHz Kryo 475 Silver), 8 GB RAM, Li-Ion battery capacity 4080 mAh and, Adreno 620 GPU. We used Pixel 5 for real-world applications. For the power supply, we use a Topward 6306D Dual Tracking DC Power Supply~\citep{power-supply}. \name\ server is a 16-core machine with 2.7GHz Intel Xeon CPU ES-2680 processors, 6 GB RAM, and running CentOS Linux 7.4.1708. 

\noindent \textbf{Workloads: }We implemented several workloads in SALSA, based on benchmarks in~\cite{IMCM-full}, utilizing different mobile devices: compute, network, memory, and I/O. We implemented various execution profiles using these workloads, and a simulator to experiment with complex profiles for \actor{}s different power states in various execution configurations.

\begin{figure}
    \centering
    \includegraphics[width=0.75\textwidth]{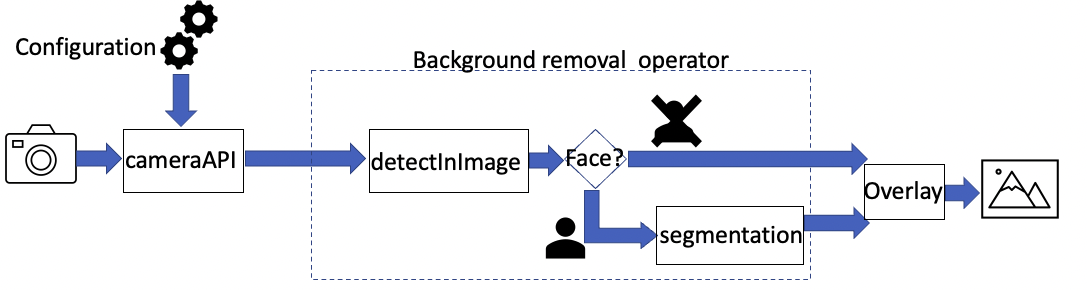}
    \caption{ \normalfont Video conferencing app which detects faces in  video frames, and conditionally executes segmentation to identify background and overlay with a different image.}
    \label{fig:escope-video-conference-app}
\end{figure}

For real-world applications, we evaluated two applications :  (1) ML Kit-based face recognition application\cite{mlkitface}, as shown in  Fig~\ref{fig:face-detection-pipeline} and, (2)  video conferencing application shown in Fig~\ref{fig:escope-video-conference-app}. The face recognizer identifies human faces and detects facial expressions (such as smiling or frowning). ML Kit face detection application extracts the video frames from the camera API and configures them to a resolution set within the application. Then the face detector (from the MLKit library) identifies a bounding box around the face in the frame if a face exists and then sends it to the facial expression classifier. 

The video conferencing application detects a face, identifies the segmentation mask of the background, and then removes the background. Such a feature has become widespread for video conferencing applications, allowing users to remove/change the background. We leverage the face detection library from the first application. After face detection, a segmentation module is executed on the frame, which provides a mask for the background to be applied to selected frames, allowing us to remove the background and overlay it with a different background image. We set the polling interval at 1 second for all our execution segments. 

\subsection{Feasibility of \name\ for mobile devices}
\label{sec:overhead-eval}

In \name, training and prediction happen alongside other workloads running on the mobile device. Hence, a critical design goal is to minimize monitoring resource overhead. We study the feasibility of deploying \name\ on a mobile device by investigating its power and CPU utilization overhead. 

\name\ client has an overhead on the mobile device associated with each of the phases shown in Figure~\ref{fig:monitoring-app}. For \textit{Data collection phase}, periodic sampling of training data incurs an additional average power cost of 225mW, while the average idle power on the mobile device is 717mW. We could reduce the collection cost by increasing the polling interval. In the \textit{Data Uploading phase}, sending a log file (average: 802.1KB) has a negligible impact on energy (\textasciitilde17.94J), which is about the cost of a single load for certain Web pages~\citep{RECON}. Finally, in the \textit{prediction phase:} After identifying and training the best model, we deploy it on the mobile device and run it periodically on the sampled execution profiles. Rest of this section describes the power and CPU utilization overhead associated with these models.   

\begin{figure}[htp]
\centering
    \includegraphics[width=0.60\textwidth]{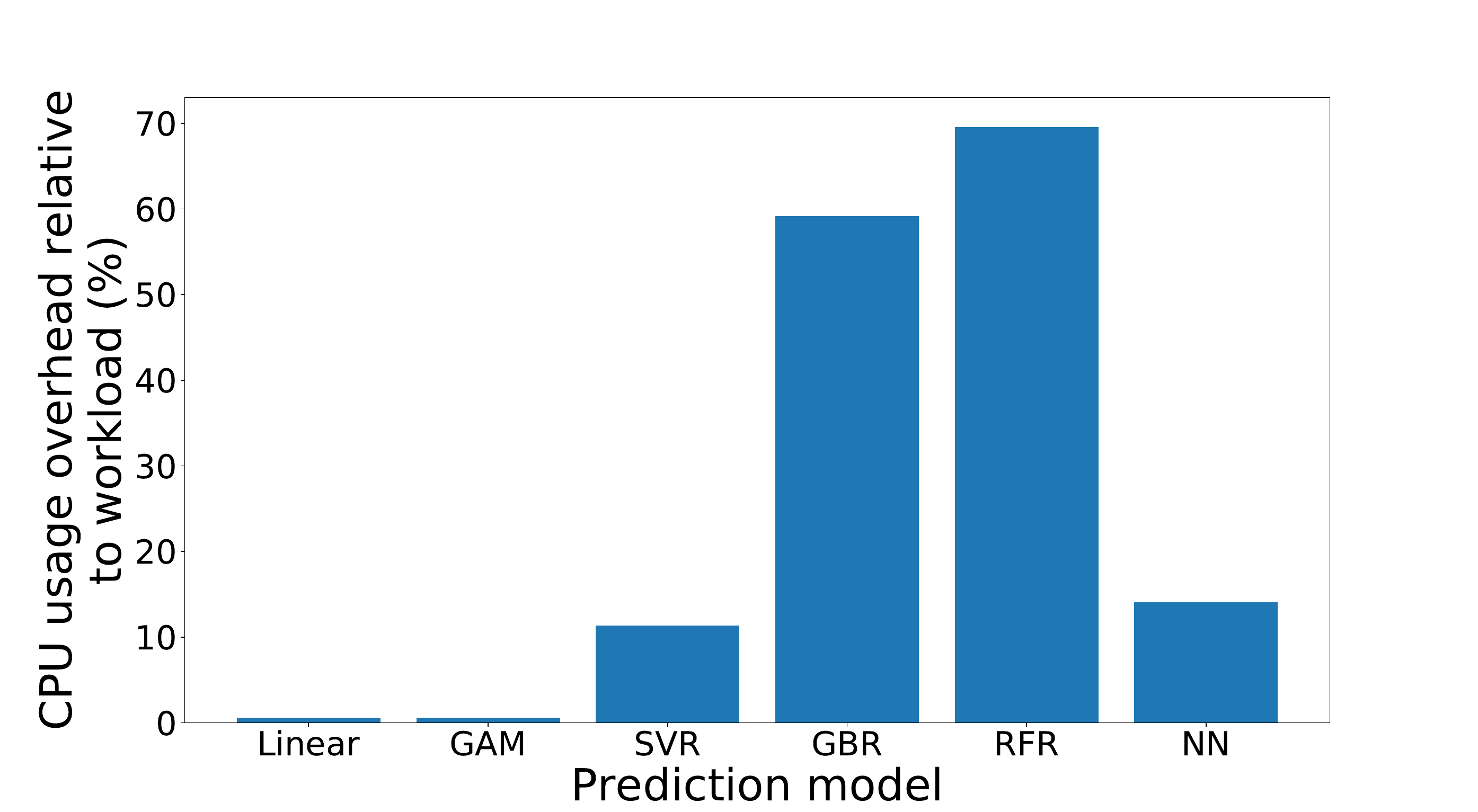}
    \caption{\normalfont CPU usage overhead of different models while training for a fixed workload}
    \label{fig:CPUOverhead}
\end{figure}

\noindent \textbf{CPU utilization overhead: } We compare the CPU utilization overhead of different prediction models because they are compute-intensive. We use a fixed NQueens workload to analyze each prediction model's overhead (Fig.~\ref{fig:CPUOverhead}). Linear models and GAMs are the least expensive owing to the simple arithmetic operations after determining the model's coefficients. Following them, SVR and NN incur moderate overhead. Finally, RFRs and GBRs use regression trees as their base learners. Though a larger number of trees make them more stable predictors, it comes at a significantly high computation cost, as seen in Fig.~\ref{fig:CPUOverhead}. 

\noindent \textbf{Power overhead:} Linear models and GAMs incur negligible power overhead. While NNs are more expensive than SVR. However, even NNs have an additional power cost of only 72.5 mW, which is significantly cheaper (\textasciitilde30\%) than the power overhead of data collection. Optimizations like caching power predictions from a previously observed data sample can further reduce the overhead.

We observed that \name\ presents low overhead compared to idle power on the mobile device. We discard RFRs \& GBRs models, due to their high overhead, making them unfeasible for deployment.
\subsection{Accuracy of power measurements}
\label{sec:energy-measure-acc}
 We rely on battery SoC estimates for measuring power draw on the mobile device (Section~\ref{sec:energy-measurement-proposed}), with a measurement error of less than 1\% \cite{batterySoC}, which directly affects our power predictions.

To verify the accuracy of battery SoC, we implemented an application to vary the screen brightness (\textit{DispApp}) providing a fixed workload. Comparing the measurement of battery SoC and ground truth using power monitor for \textit{DispApp}, we observed a measurement error of 0.59\%, corroborating the observation of previous work~\cite{batterySoC}.

\subsection{Accuracy of \name{} }
\label{sec:perf-eval}
We evaluated \name\ across different execution profiles to show its high confidence of power prediction.

\subsubsection{Fixed workloads}
\begin{figure}[htp]
\centering
\includegraphics[width=0.85\textwidth]{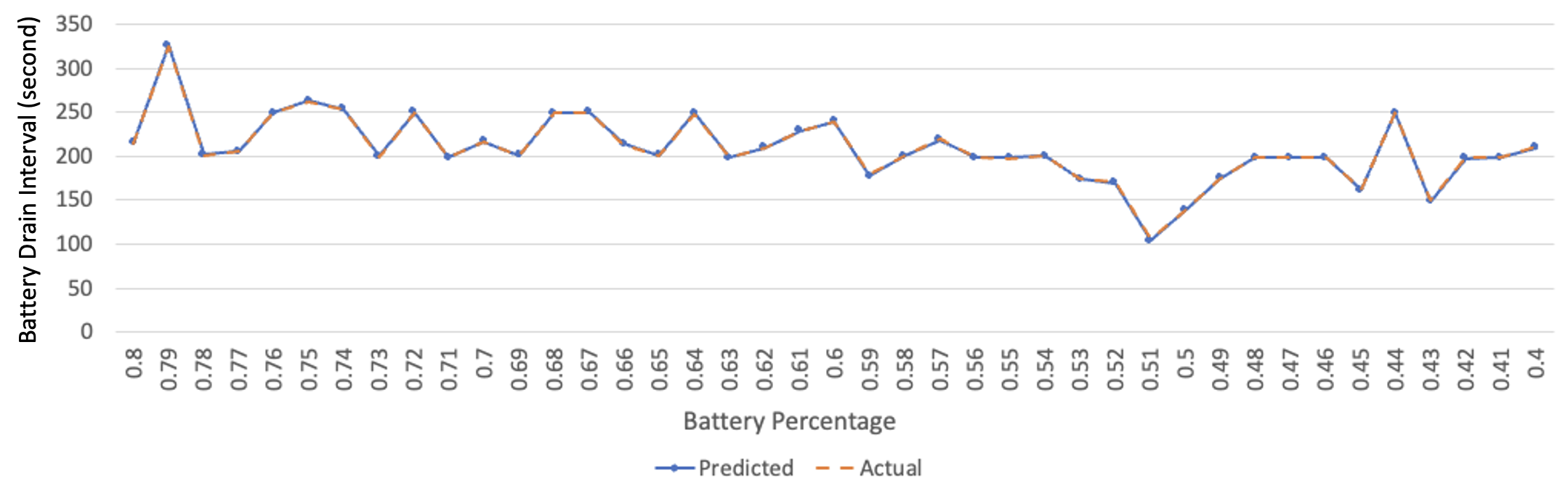}
    \caption{\normalfont  \name{} predicts the battery drain interval for all synthetic fixed workload with very high accuracy ( > 95\%) using the linear model. This figure shows the performance of \name{} for a memory intensive (Fibonacci) workload  }
    \label{fig:FixedWkldPred}
\end{figure}

: Periodic tasks are common in mobile applications, e.g., manual animation, uploading data over a network, etc. To evaluate \name\ for such tasks, we implemented fixed workload profile with a fixed period and duration.
\begin{table*}
  \begin{tabular}[width=\linewidth]{p{0.11\linewidth} p{0.06\linewidth} p{0.30\linewidth} p{0.06\linewidth} p{0.06\linewidth} p{0.06\linewidth} p{0.20\linewidth} } 
    \toprule
     \textbf{Dominant Resource} & \textbf{Load} & \textbf{Description} & \textbf{Linear RMSE} & \textbf{SVR RMSE} & \textbf{NN RMSE} & \textbf{Predicted power(Watt)$^\dagger$ (Accuracy$ \ddagger $ )} \\
    \midrule
    \multirow{3}{*}{\textbf{Compute}} & \textbf{NQ} & Places N Queens on N*N chess board & 15.56 & 13.76 & 102.8 & 3.2W(97\%) \\ 			\cline{2-7}
               & \textbf{Trap} & Area under the curve using trapezoidal rule & 14.85 & 16.82 & 55.56 & 1.4W(98.6\%)\\ \cline{2-7}
            & \textbf{Nums} & Concurrent arithmetic operations & 17.65 & 16.82 & 57.93 & 1.7W(98.1\%)\\
			\hline
           \textbf{Memory} & \textbf{Fib} & Calculates Fibonacci series  & 11.9 & 11.63 & 88.52 & 1.8W(98.6\%)\\
			\hline
    \multirow{4}{*}{\parbox{0.11\linewidth}{\centering \textbf{Compute +I/O}}} & \textbf{Sort} & External sort of contents of a file & 4.32 & 5.63 & 71.99 & 4.4W(98.8\%) \\ \cline{2-7}
    & \textbf{Image} & Read image and extract Haar-like features~\cite{haar} & 9.92 & 9.94 & 70.03 & 0.9W(99.4\%)\\ \cline{2-7}
    & \textbf{NQ+} & \multirow{2}{*}{  Concurrently run NQ \& Sort} & \multirow{2}{*}{12.35} & \multirow{2}{*}{9.5} & \multirow{2}{*}{50.69} & NQ:2.7W\\
                & \textbf{Sort}& & & & & Sort:4.0W(97.3\%)\\\hline
        
            \textbf{ Net+ I/O} & \textbf{Scp} & Copy file from mobile to server & 12.32 & 12.3 & 122.86 & 0.52W(99.5\%)\\
			\hline

 			 \textbf{Compute+} & \multirow{2}{*} {\textbf{NQ+Fib}} & \multirow{2}{*} { Concurrently run NQ \& Fib} & \multirow{2}{*} {13.87} & \multirow{2}{*} {16.58} & \multirow{2}{*} {90.38} & NQ:3.5W\\
            \textbf{Memory} & &  & & & & Fib:2.5W (97.3\%)\\
			\hline

              \textbf{Compute+}  &   \multirow{2}{*}{\textbf{NQ+Scp}} & \multirow{2}{*}{ Concurrently run NQ \& Scp} & \multirow{2}{*}{11.2} & \multirow{2}{*}{10.72} & \multirow{2}{*}{85.63} & NQ:2.8W\\
              \textbf{Net+ I/O} & & & & & & Scp:0.9W(98.6\%)\\

  \bottomrule
\end{tabular}
\caption{\normalfont Results from fixed load benchmarks. $^\dagger$ Using the linear model.$ \ddagger $ Accuracy of the battery drain interval prediction.
    }\label{tab:fixed-workloads}
\end{table*}
Table ~\ref{tab:fixed-workloads} shows the relative power of \actor{}s and the accuracy of various estimation models such as linear, SVR, and NN. For accuracy, we use average RMSE using an 8-fold nested CV.  Linear models and SVRs, are very good at estimating drain interval lengths based on \actor{} execution profiles. They have comparable accuracies, with SVR being slightly better (\textasciitilde2.1\%). 
Fig\ref{fig:FixedWkldPred} shows that linear models can accurately predict the battery drop intervals while executing fixed workloads. 
Hence, linear models are preferred for fixed workloads as \actor{}s execute in a fixed power state, given their low resource overhead and high accuracy.

NNs perform poorly on our workloads. Training NNs is difficult due to the large hyperparameter search space, and poor accuracy due to the lack of large training data. We validated the impact of training data size by merging data from five runs of battery discharge, and observed improved accuracy with increased data size. We exclude GAMs from our analysis, as they require manual parameter tuning from each application, significantly increasing variance in accuracy across applications and reducing scalability.

\subsubsection{Variable workloads.} Mobile applications also contain tasks without a fixed duration/period, such as interactive tasks whose run time depends on user input. To model such applications and evaluate the impact of changes in workload on battery SoC measurements across discharge intervals, we run variable compute and network intensive workloads for different durations across discharge intervals.\\
\textit{Compute:} We implemented a compute-intensive \actor{}, which creates 6 \actor{} instances generating random numbers and performing arithmetic operations on them, scheduled to alternate between 120 and 10 seconds across discharge intervals. \name\ predicted the  power of \actor{} to be 0.84W$\pm$0.05W at 95\% confidence.\\
\textit{Network:} We implemented network-intensive $Upload$ \actor{}, which uploads 11.29 MB of data every second using WiFi with upload bandwidth of 115 Mbps. We maintain the data in memory, which removes I/O cost, unlike $scp$. This application alternates between 120 and 10 seconds across discharge intervals. \name\ predicts the  power of this \actor{} to be 0.91W$\pm$ 0.1W at 95\% confidence.

\subsubsection{Real-World Applications}
\label{sec:escope-real-world-apps}

\begin{figure}[htp]
\centering
\includegraphics[width=0.95\textwidth]{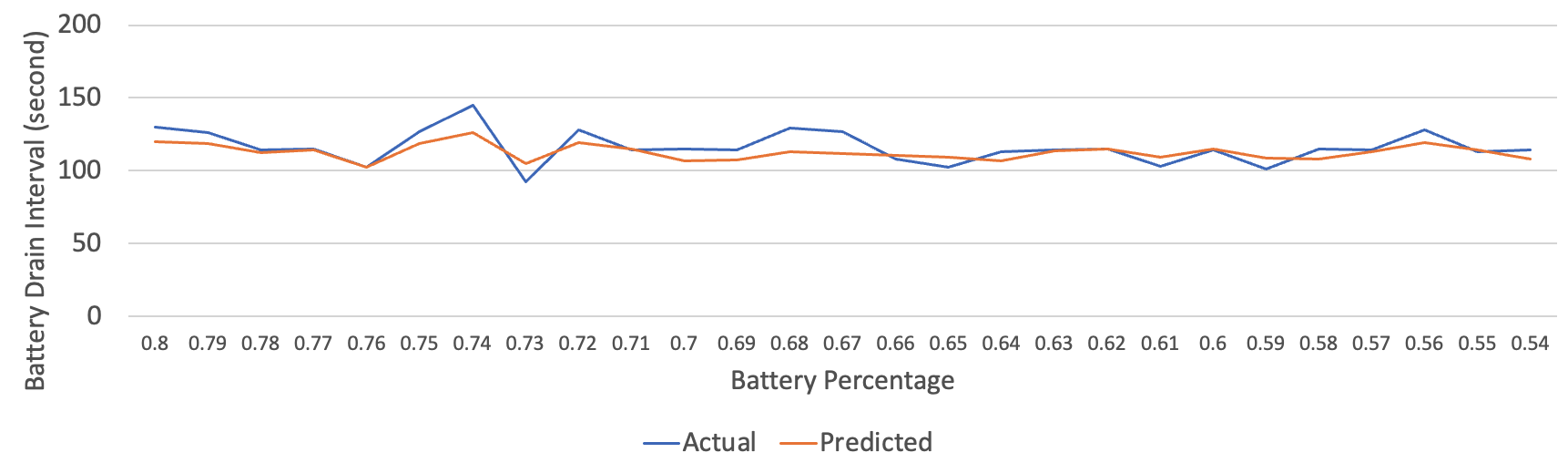}
    \caption{\normalfont Predicting the battery drain interval while the face-recognizer application is executing over multiple battery interval drops }
    \label{fig:BatPred}
\end{figure}

We evaluated two real-world applications, \textit{face recognizer} \cite{mlkitface} and \textit{video conference} (Fig. ~\ref{fig:escope-video-conference-app}), with a variety of inputs and configurations. We configured the resolution and frame rate of the video frames by sweeping through resolution values between 144p to 1080p,  and sampling rate in application. This provides insight into how these factors impact power draw. Previous works ~\cite{chameleon-video-analytics} have observed that frame rate and resolution significantly impact video analytics applications' energy.
To study the input dependency of \actor{} power consumption, we varied the input frame content as follows. First, we input images \textit{without faces}:  by pointing the camera toward an empty background (simulating the case of a human user away from the phone). Then, we set the input \textit{with faces}: by pointing the camera towards a screen containing a human face covering 50\% of the screen area. We observed that overlaying background image consumes negligible power compared to other \actor{}s in the video conference application, hence did not consider it to simplify our analysis.
\newline
\noindent \textit{Face recognizer:} The linear regressor model of \name{} predicted the workload's battery discharge interval duration with an RMSE of  17.6$\pm$11.24 seconds which corresponds to 0.24W (3\%) error in power prediction. The SVR model has an RMSE of 14.3$\pm$9 seconds (0.12W (1.5\%) error in power prediction). Despite the higher overhead, we deployed the SVR model for inference on the mobile device due to higher accuracy than the linear model, as observed in Fig \ref{fig:BatPred}.

\begin{figure}[htp]
\centering
\includegraphics[width=0.95\textwidth]{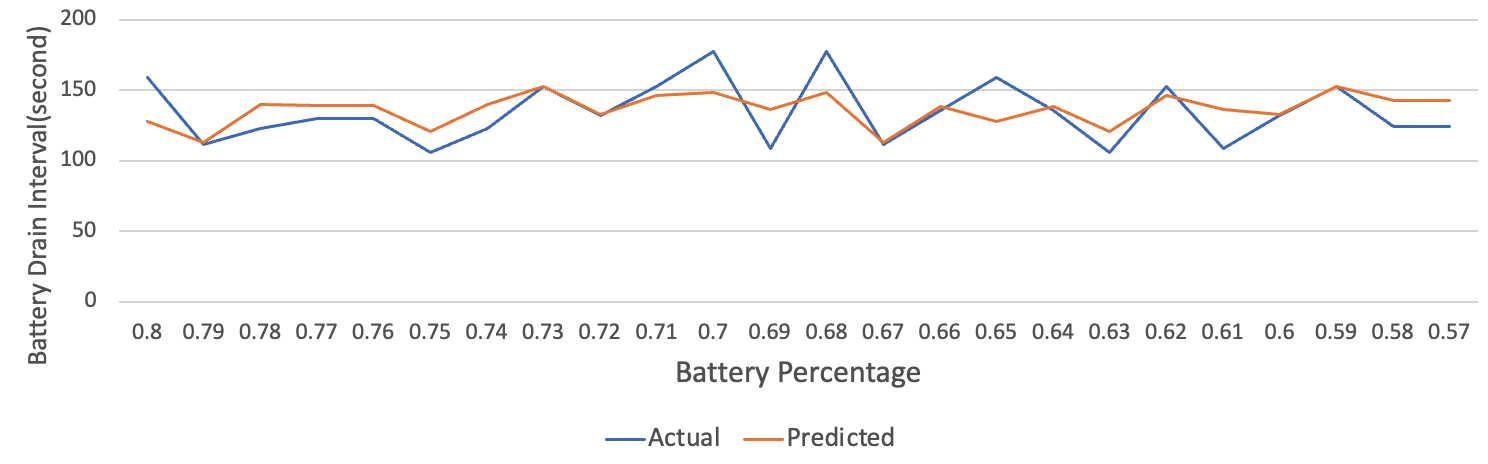}
    \caption{\normalfont Predicting the battery drain interval while the video-conference application is executing over multiple battery interval drops }
    \label{fig:BatPow}
\end{figure}

\noindent \textit{Video conference application:} We observe an RMSE of 26.37 $\pm$ 12.9 seconds (0.23W (3.4\%) error in power prediction) for the linear regressor model, while SVR gave an RMSE of 24.7$\pm$16.5 (0.25W (3.6\%) error in power prediction) in the battery discharge interval duration. The difference in accuracy between the two models is not significant. Hence \name{} chooses the linear model for inference (Fig \ref{fig:BatPow}) for this application.

\name{} uses a heuristic to choose a training model, where we choose SVR if it outperforms the linear model by $>$15\% in discharge interval prediction accuracy, despite the higher overhead. We observed that accuracy could be improved by collecting more execution logs on the mobile device during the training phase. Thus, more powerful models such as SVR can enhance the power prediction accuracy for complex real-world applications.

Based on different models' accuracy and overhead, we identify linear and SVR as ideal candidates for training on \name\ server. Linear models have minimal resource overhead, while SVR can better handle complex execution profiles (details in \S~\ref{sec:non-linear-wkld}) with an acceptable increase in overhead.

\subsubsection{Comparison with related work}
We compare \name\ with MAUI and RECON, based on our discussion in \S\ref{sec:related-work}. MAUI~\cite{cuervo2010maui} is a code offloading framework that relies on profiling energy of each resource and instrumenting source code to estimate resource usage of each application component. While RECON~\cite{RECON} monitors both the resource usage and application components for a fine-grained application power modeling.
\begin{figure}[htp]
\centering
\includegraphics[width=0.75\textwidth]{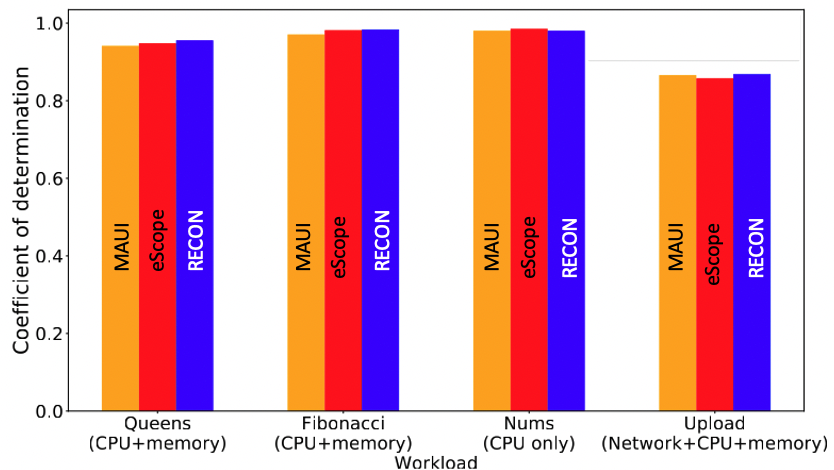}
    \caption{\normalfont The prediction accuracy (coefficient of determination) of \name\ is similar to that of MAUI and RECON}
    \label{fig:relwork-rsquared-resusage}
\end{figure}
 

We compared our performance against the prior works on the synthetic workloads utilizing different resources ( compute, memory, and network). MAUI requires code annotations to measure the fine grained \actor{}-level power values, which is readily available for these synthetic workloads. Additionally, since both MAUI and RECON requires hardware model, these synthetic benchmarks have a diverse resource usage profile, allowing us to compare the accuracy of our model in different scenario. We observed that \name{}'s  prediction  accuracy is within 99\%  of MAUI and RECON (Fig. ~\ref{fig:relwork-rsquared-resusage} ). We used fine-grained \actor{} execution traces which closely correspond to resource utilization changes during execution that impacts power consumption. For complex workloads where the \actor{}s may execute in significantly different power states when scheduled, \name\ relies on non-linear models such as SVR to accurately model the \actor{}s power behavior.
We should note that, RECON might perform better for bursty short living workload (for e.g., webpage loading), as the short term behaviour might be better captured by the resource model.

\subsection{ Summary for Prediction Models}
\label{sec:Summary}

We summarize our analysis and insights based on the different prediction models that were studied in our evaluation:
\begin{enumerate}[label=\textbullet,leftmargin=0.3cm]
\item \textbf{Linear models} have low overhead and high accuracy  (<17.65 seconds error in estimating drain interval) for all evaluated applications (benchmark and real-world). Linear models perform well if the observation interval (execution segment) granularity is finer than the prediction interval (drain interval). As long as the complexity in energy consumption behavior is contained within a drain interval, they do not affect model accuracy.

\item \textbf{SVRs} have acceptable overhead, while their accuracy is slightly better than linear models. It can capture non-linearities and work incredibly well for complex datasets with a large number of features and small training observations. 

\end{enumerate}

We considered other models, such as  \textbf{GAM}, \textbf{RFRs, GBRs}, and \textbf{NN}, but they were unsuitable due to high overhead (RFRs and GBRs) or low accuracy(GAMs, NNs). SVRs are accurate and have acceptable overhead, though their CPU overhead is higher than linear models at \textasciitilde10\%, the higher overhead of SVRs would be acceptable if they can better capture complex execution profiles. Lasso regression has its limitations (see ~\cite{lassolimit}), as it suffers if \actor{}s execute in multiple power states or there are dependencies between features in training data. Each feature in the training data corresponds to an \actor{} type and its instance count.

\subsection{Detailed Analysis of \Actor{}s in Multiple Power States}
\label{sec:non-linear-wkld}
\name{} benefits from complex prediction models, such as SVR, for real-world (\S~\ref{sec:escope-real-world-apps}) and other mobile applications exhibiting complex resource usage characteristics. For e.g., loading web pages containing components utilizing different resources, such as compute and then network in an image/gif download component for \textit{fico.com}~\cite{RECON}. We simulate complex execution profiles to investigate non-linear power characteristics between segment power and execution duration and benefits of using SVR.

\begin{figure}
\begin{minipage}[b]{0.52\textwidth}
 \includegraphics[width=\textwidth]{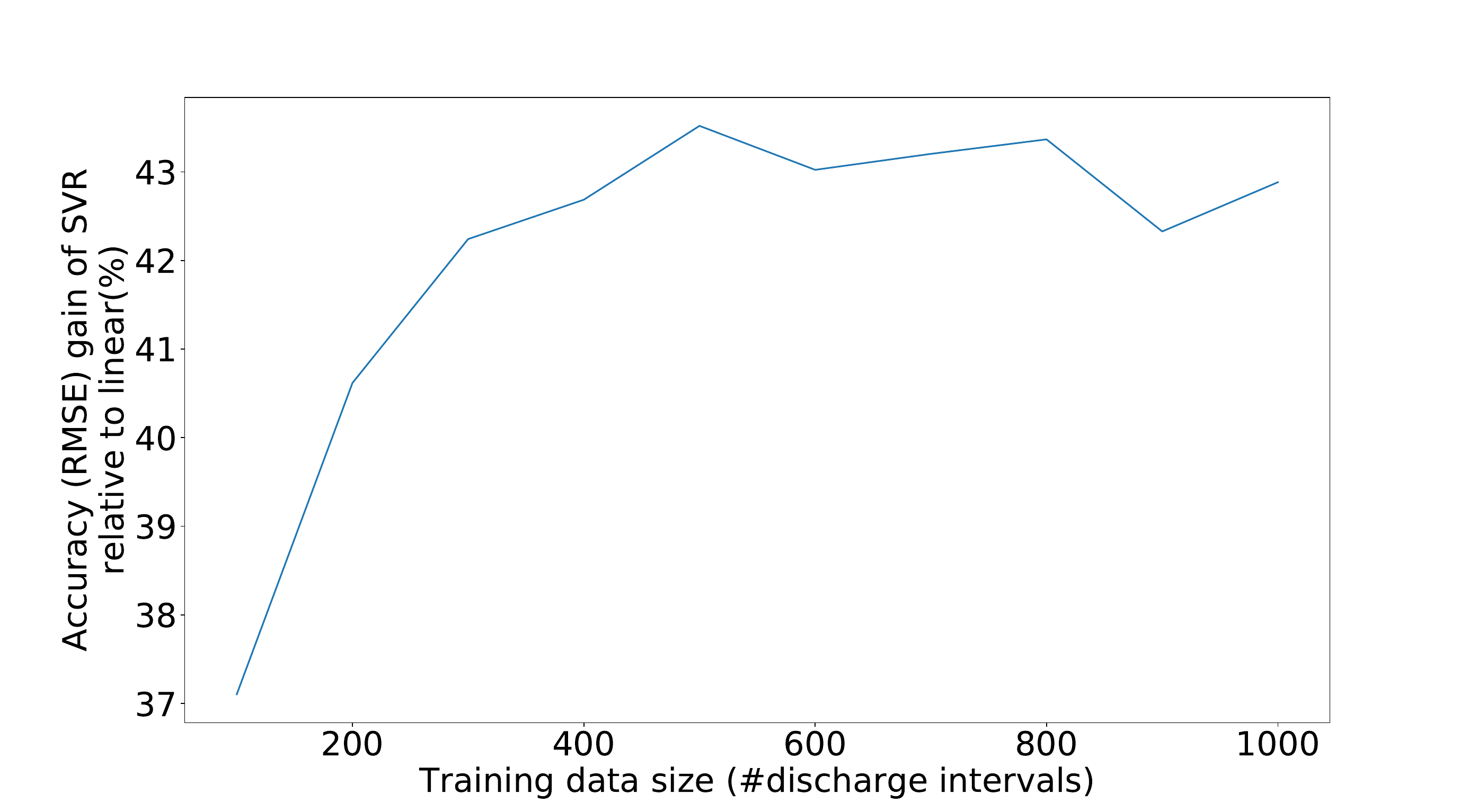}\\
\subcaption{\normalfont Accuracy gain of SVR relative to linear model\\ with increasing training data size.}
\label{fig:nonlinearaccuracyvstrainingsize}
\end{minipage}%
\begin{minipage}[b]{0.52\textwidth}
\includegraphics[width=\textwidth]{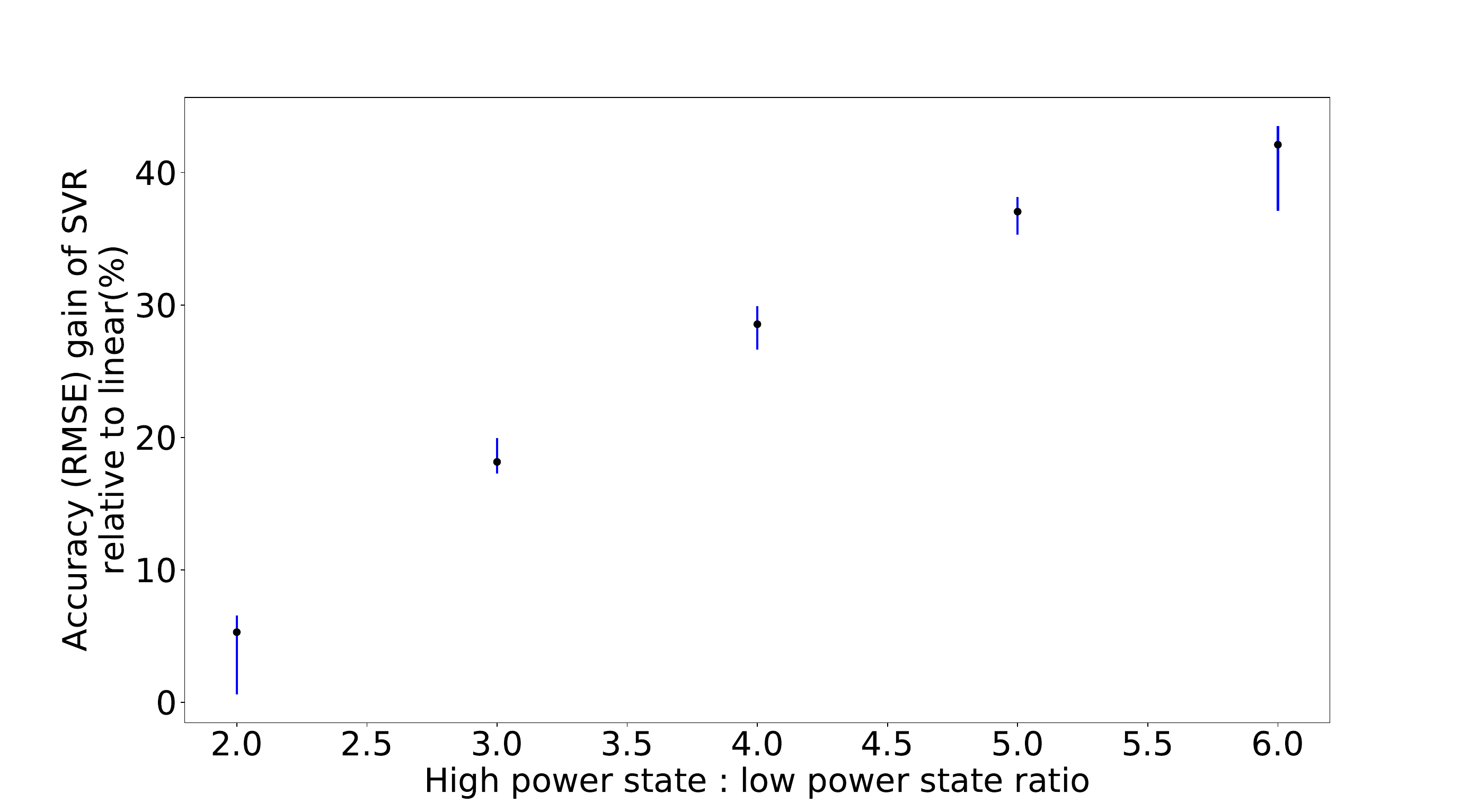}\\
\subcaption{\normalfont  Accuracy gain of SVR relative to linear model \\ with increasing power ratio between high and low \\ power state.}
    \label{fig:nonlinearaccuracyvspowerratio}
\end{minipage}%
 \caption{\normalfont Comparing the accuracy of SVR and Linear models in \name{} }
\end{figure}

The simulator generates different execution configurations, while using real world datasets we had collected from mobile devices to model discharge intervals, active \actor{}s in execution segments and noise characteristics. Figure~\ref{fig:nonlinearaccuracyvspowerratio} shows that SVR can perform significantly better than linear models (up to 40\%) for reasonably large ratios between \actor{} power states. For a fixed power state ratio of an \actor{}, Figure~\ref{fig:nonlinearaccuracyvstrainingsize} shows the trend of increasing accuracy gain of SVR relative to linear, with increase in the training data size. We observed a similar trend for other power ratios we evaluated as well. So we see that linear models have low resource overhead but SVR can provide significant accuracy gains when monitoring non-linear workloads.

 \section{Conclusion \& Future Work}
\label{sec:conclusion}

This paper proposed and implemented \name{}, a power prediction technique for estimating the energy costs of video analytics applications on mobile devices. \name{}  predicts operator-level power consumption using device-level energy draw information in conjunction with an application's execution traces. Additionally, it does not use any resource-usage-based power model. \name{} is highly scalable and portable, as the prediction models are trained on a server using device-specific data independently. Thus modeling for new devices or retraining requires additional computing on a server without any data from previous training, reducing data cost. Our approach meets the stated design goal of being fine-grained, resource-agnostic, adaptable (does not require code-level changes), dynamic (captures the impact of changing input), and easily portable to new hardware and applications. Our evaluation showed that \name{} could accurately model a diverse set of workloads with minimal overhead. We also evaluated several ML models to analyze their feasibility for our approach. \name{} can be integrated with a mobile-cloud placement framework that optimizes mobile energy consumption. We can facilitate such a placement framework by identifying energy-intensive operators, which enables the framework to place them on a remote server, reducing on-device computational power costs. In the future, we plan to support a wider variety of languages, provide support beyond mobile devices (e.g., drone controllers) and investigate performance and security policies that affect power-optimization decisions.

\bibliographystyle{ACM-Reference-Format}
\bibliography{ucc_2014}


\begin{thebibliography}{70}


\ifx \showCODEN    \undefined \def \showCODEN     #1{\unskip}     \fi
\ifx \showDOI      \undefined \def \showDOI       #1{#1}\fi
\ifx \showISBNx    \undefined \def \showISBNx     #1{\unskip}     \fi
\ifx \showISBNxiii \undefined \def \showISBNxiii  #1{\unskip}     \fi
\ifx \showISSN     \undefined \def \showISSN      #1{\unskip}     \fi
\ifx \showLCCN     \undefined \def \showLCCN      #1{\unskip}     \fi
\ifx \shownote     \undefined \def \shownote      #1{#1}          \fi
\ifx \showarticletitle \undefined \def \showarticletitle #1{#1}   \fi
\ifx \showURL      \undefined \def \showURL       {\relax}        \fi
\providecommand\bibfield[2]{#2}
\providecommand\bibinfo[2]{#2}
\providecommand\natexlab[1]{#1}
\providecommand\showeprint[2][]{arXiv:#2}

\bibitem[pow(1995)]%
        {power-supply}
 \bibinfo{year}{1995}\natexlab{}.
\newblock \bibinfo{title}{Topward 6306D Dual Tracking DC Power Supply 0-30V/0-6A x 2; 5V/5A}.
\newblock \bibinfo{howpublished}{\url{https://www.topward.com.tw/data/6000.pdf}}.
\newblock


\bibitem[and(2008)]%
        {android}
 \bibinfo{year}{2008}\natexlab{}.
\newblock \bibinfo{title}{Android}.
\newblock \bibinfo{howpublished}{\url{https://www.android.com/}}.
\newblock


\bibitem[And(2008)]%
        {AndroidBatteryManager}
 \bibinfo{year}{2008}\natexlab{}.
\newblock \bibinfo{title}{Android Battery Manager Service}.
\newblock \bibinfo{howpublished}{\url{https://developer.android.com/reference/android/os/BatteryManager}}.
\newblock


\bibitem[mon(2014)]%
        {monsoonpowermonitor}
 \bibinfo{year}{2014}\natexlab{}.
\newblock \bibinfo{title}{Monsoon Power Monitor.}
\newblock \bibinfo{howpublished}{\url{https://www.msoon.com/LabEquipment/PowerMonitor/}}.
\newblock


\bibitem[kot(2016)]%
        {kotlin-actors}
 \bibinfo{year}{2016}\natexlab{}.
\newblock \bibinfo{title}{Actors in Kotlin}.
\newblock \bibinfo{howpublished}{\url{https://github.com/Kotlin/kotlinx.coroutines/blob/master/docs/shared-mutable-state-and-concurrency.md\#actors}}.
\newblock


\bibitem[Sal(2016)]%
        {SalsaOnAndroid}
 \bibinfo{year}{2016}\natexlab{}.
\newblock \bibinfo{title}{SALSA on Android}.
\newblock \bibinfo{howpublished}{\url{http://osl.cs.illinois.edu/software/salsa-android/}}.
\newblock


\bibitem[2014.(2014)]%
        {maximdatasheet}
\bibfield{author}{\bibinfo{person}{Maxim. 2014.}} \bibinfo{year}{2014}\natexlab{}.
\newblock \bibinfo{booktitle}{\emph{{MAX17047/MAX17050, ModelGauge m3 Fuel Gauge.}}}
\newblock \bibinfo{type}{{T}echnical {R}eport}.
\newblock


\bibitem[Abadi et~al\mbox{.}(2015)]%
        {tensorflow}
\bibfield{author}{\bibinfo{person}{Martín Abadi}, \bibinfo{person}{Ashish Agarwal}, \bibinfo{person}{Paul Barham}, \bibinfo{person}{Eugene Brevdo}, \bibinfo{person}{Zhifeng Chen}, \bibinfo{person}{Craig Citro}, \bibinfo{person}{Greg Corrado}, \bibinfo{person}{Andy Davis}, \bibinfo{person}{Jeffrey Dean}, \bibinfo{person}{Matthieu Devin}, \bibinfo{person}{Sanjay Ghemawat}, \bibinfo{person}{Ian Goodfellow}, \bibinfo{person}{Andrew Harp}, \bibinfo{person}{Geoffrey Irving}, \bibinfo{person}{Michael Isard}, \bibinfo{person}{Yangqing Jia}, \bibinfo{person}{Rafal Jozefowicz}, \bibinfo{person}{Lukasz Kaiser}, \bibinfo{person}{Manjunath Kudlur}, \bibinfo{person}{Josh Levenberg}, \bibinfo{person}{Dan Mané}, \bibinfo{person}{Rajat Monga}, \bibinfo{person}{Sherry Moore}, \bibinfo{person}{Derek Murray}, \bibinfo{person}{Chris Olah}, \bibinfo{person}{Mike Schuster}, \bibinfo{person}{Jonathon Shlens}, \bibinfo{person}{Benoit Steiner}, \bibinfo{person}{Ilya Sutskever}, \bibinfo{person}{Kunal Talwar}, \bibinfo{person}{Paul
  Tucker}, \bibinfo{person}{Vincent Vanhoucke}, \bibinfo{person}{Vijay Vasudevan}, \bibinfo{person}{Fernanda Viégas}, \bibinfo{person}{Oriol Vinyals}, \bibinfo{person}{Pete Warden}, \bibinfo{person}{Martin Wattenberg}, \bibinfo{person}{Martin Wicke}, \bibinfo{person}{Yuan Yu}, {and} \bibinfo{person}{Xiaoqiang Zheng}.} \bibinfo{year}{2015}\natexlab{}.
\newblock \bibinfo{title}{TensorFlow: Large-Scale Machine Learning on Heterogeneous Distributed Systems}.
\newblock
\newblock
\urldef\tempurl%
\url{http://download.tensorflow.org/paper/whitepaper2015.pdf}
\showURL{%
\tempurl}


\bibitem[Agha et~al\mbox{.}(2018)]%
        {ACCBook}
\bibfield{author}{\bibinfo{person}{Gul Agha}, \bibinfo{person}{Minas Charalambides}, \bibinfo{person}{Kirill Mechitov}, \bibinfo{person}{Karl Palmskog}, \bibinfo{person}{Atul Sandur}, {and} \bibinfo{person}{Reza Shiftehfar}.} \bibinfo{year}{2018}\natexlab{}.
\newblock \showarticletitle{Inferring and Enforcing Use Patterns for Mobile Cloud Assurance}.
\newblock In \bibinfo{booktitle}{\emph{Assured Cloud Computing}}, \bibfield{editor}{\bibinfo{person}{R.~H. Campbell} {and} \bibinfo{person}{K.~A.~Kwiat C.~A.~Kamhoua}} (Eds.). \bibinfo{publisher}{Wiley/IEEE Press}, \bibinfo{pages}{237--376}.
\newblock


\bibitem[Agha et~al\mbox{.}(2022)]%
        {self}
\bibfield{author}{\bibinfo{person}{Gul Agha}, \bibinfo{person}{Dipayan Mukherjee}, {and} \bibinfo{person}{Atul Sandur}.} \bibinfo{year}{2022}\natexlab{}.
\newblock \showarticletitle{Performance, Energy and Parallelism: Using Near Data Processing in Utility and Cloud Computing}. In \bibinfo{booktitle}{\emph{2022 IEEE/ACM 15th International Conference on Utility and Cloud Computing (UCC)}}. \bibinfo{pages}{173--180}.
\newblock
\urldef\tempurl%
\url{https://doi.org/10.1109/UCC56403.2022.00031}
\showDOI{\tempurl}


\bibitem[{Android}(2021)]%
        {battCheck}
\bibfield{author}{\bibinfo{person}{{Android}}.} \bibinfo{year}{2021}\natexlab{}.
\newblock \bibinfo{title}{{Profile battery usage with Batterystats and Battery Historian}}.
\newblock \bibinfo{howpublished}{\url{https://developer.android.com/topic/performance/power/setup-battery-historian}}.
\newblock


\bibitem[{Apple}(2019)]%
        {iosIntrument}
\bibfield{author}{\bibinfo{person}{{Apple}}.} \bibinfo{year}{2019}\natexlab{}.
\newblock \bibinfo{title}{{Measure the energy impact of an iOS device}}.
\newblock \bibinfo{howpublished}{\url{https://web.archive.org/web/20200620073428/https://help.apple.com/instruments/mac/current/\#/deva0db8947}}.
\newblock


\bibitem[Bellosa(2000)]%
        {JouleWatcher}
\bibfield{author}{\bibinfo{person}{Frank Bellosa}.} \bibinfo{year}{2000}\natexlab{}.
\newblock \showarticletitle{The Benefits of Event-Driven Energy Accounting in Power-Sensitive Systems}. In \bibinfo{booktitle}{\emph{Proceedings of the 9th ACM SIGOPS European Workshop}}. \bibinfo{address}{Kolding, Denmark}.
\newblock
\urldef\tempurl%
\url{http://i30www.ira.uka.de/research/publications/}
\showURL{%
\tempurl}


\bibitem[Breiman(2001)]%
        {RFR}
\bibfield{author}{\bibinfo{person}{Leo Breiman}.} \bibinfo{year}{2001}\natexlab{}.
\newblock \showarticletitle{Random Forests}.
\newblock \bibinfo{journal}{\emph{Machine Learning}} \bibinfo{volume}{45}, \bibinfo{number}{1} (\bibinfo{date}{01 Oct} \bibinfo{year}{2001}), \bibinfo{pages}{5--32}.
\newblock
\showISSN{1573-0565}
\urldef\tempurl%
\url{https://doi.org/10.1023/A:1010933404324}
\showDOI{\tempurl}


\bibitem[Bridges et~al\mbox{.}(2016)]%
        {gpupowermodelsurvey}
\bibfield{author}{\bibinfo{person}{Robert~A. Bridges}, \bibinfo{person}{Neena Imam}, {and} \bibinfo{person}{Tiffany~M. Mintz}.} \bibinfo{year}{2016}\natexlab{}.
\newblock \showarticletitle{Understanding GPU Power: A Survey of Profiling, Modeling, and Simulation Methods}.
\newblock \bibinfo{journal}{\emph{ACM Comput. Surv.}} \bibinfo{volume}{49}, \bibinfo{number}{3}, Article \bibinfo{articleno}{41} (\bibinfo{date}{sep} \bibinfo{year}{2016}), \bibinfo{numpages}{27}~pages.
\newblock
\showISSN{0360-0300}
\urldef\tempurl%
\url{https://doi.org/10.1145/2962131}
\showDOI{\tempurl}


\bibitem[Cai et~al\mbox{.}(2017)]%
        {neuralpower}
\bibfield{author}{\bibinfo{person}{Ermao Cai}, \bibinfo{person}{Da{-}Cheng Juan}, \bibinfo{person}{Dimitrios Stamoulis}, {and} \bibinfo{person}{Diana Marculescu}.} \bibinfo{year}{2017}\natexlab{}.
\newblock \showarticletitle{NeuralPower: Predict and Deploy Energy-Efficient Convolutional Neural Networks}. In \bibinfo{booktitle}{\emph{Proceedings of The 9th Asian Conference on Machine Learning, {ACML} 2017, Seoul, Korea, November 15-17, 2017}} \emph{(\bibinfo{series}{Proceedings of Machine Learning Research}, Vol.~\bibinfo{volume}{77})}, \bibfield{editor}{\bibinfo{person}{Min{-}Ling Zhang} {and} \bibinfo{person}{Yung{-}Kyun Noh}} (Eds.). \bibinfo{publisher}{{PMLR}}, \bibinfo{pages}{622--637}.
\newblock
\urldef\tempurl%
\url{http://proceedings.mlr.press/v77/cai17a.html}
\showURL{%
\tempurl}


\bibitem[Cao et~al\mbox{.}(2017)]%
        {RECON}
\bibfield{author}{\bibinfo{person}{Yi Cao}, \bibinfo{person}{Javad Nejati}, \bibinfo{person}{Muhammad Wajahat}, \bibinfo{person}{Aruna Balasubramanian}, {and} \bibinfo{person}{Anshul Gandhi}.} \bibinfo{year}{2017}\natexlab{}.
\newblock \showarticletitle{Deconstructing the Energy Consumption of the Mobile Page Load}.
\newblock \bibinfo{journal}{\emph{Proc. ACM Meas. Anal. Comput. Syst.}} \bibinfo{volume}{1}, \bibinfo{number}{1}, Article \bibinfo{articleno}{6} (\bibinfo{date}{June} \bibinfo{year}{2017}), \bibinfo{numpages}{25}~pages.
\newblock
\showISSN{2476-1249}
\urldef\tempurl%
\url{https://doi.org/10.1145/3084443}
\showDOI{\tempurl}


\bibitem[Cawley and Talbot(2010)]%
        {NestedCV}
\bibfield{author}{\bibinfo{person}{Gavin~C. Cawley} {and} \bibinfo{person}{Nicola~L.C. Talbot}.} \bibinfo{year}{2010}\natexlab{}.
\newblock \showarticletitle{On Over-fitting in Model Selection and Subsequent Selection Bias in Performance Evaluation}.
\newblock \bibinfo{journal}{\emph{J. Mach. Learn. Res.}}  \bibinfo{volume}{11} (\bibinfo{date}{Aug.} \bibinfo{year}{2010}), \bibinfo{pages}{2079--2107}.
\newblock
\showISSN{1532-4435}
\urldef\tempurl%
\url{http://dl.acm.org/citation.cfm?id=1756006.1859921}
\showURL{%
\tempurl}


\bibitem[Chen et~al\mbox{.}(2018)]%
        {MARVEL}
\bibfield{author}{\bibinfo{person}{Kaifei Chen}, \bibinfo{person}{Tong Li}, \bibinfo{person}{Hyung-Sin Kim}, \bibinfo{person}{David~E. Culler}, {and} \bibinfo{person}{Randy~H. Katz}.} \bibinfo{year}{2018}\natexlab{}.
\newblock \showarticletitle{MARVEL: Enabling Mobile Augmented Reality with Low Energy and Low Latency}. In \bibinfo{booktitle}{\emph{Proceedings of the 16th ACM Conference on Embedded Networked Sensor Systems}} (Shenzhen, China) \emph{(\bibinfo{series}{SenSys '18})}. \bibinfo{publisher}{Association for Computing Machinery}, \bibinfo{address}{New York, NY, USA}, \bibinfo{pages}{292–304}.
\newblock
\showISBNx{9781450359528}
\urldef\tempurl%
\url{https://doi.org/10.1145/3274783.3274834}
\showDOI{\tempurl}


\bibitem[Chen et~al\mbox{.}(2015)]%
        {glimpse-continuous-face-track-data}
\bibfield{author}{\bibinfo{person}{Tiffany Yu-Han Chen}, \bibinfo{person}{Lenin Ravindranath}, \bibinfo{person}{Shuo Deng}, \bibinfo{person}{Paramvir Bahl}, {and} \bibinfo{person}{Hari Balakrishnan}.} \bibinfo{year}{2015}\natexlab{}.
\newblock \showarticletitle{Glimpse: Continuous, Real-Time Object Recognition on Mobile Devices}. In \bibinfo{booktitle}{\emph{Proceedings of the 13th ACM Conference on Embedded Networked Sensor Systems}} (Seoul, South Korea) \emph{(\bibinfo{series}{SenSys '15})}. \bibinfo{publisher}{Association for Computing Machinery}, \bibinfo{address}{New York, NY, USA}, \bibinfo{pages}{155–168}.
\newblock
\showISBNx{9781450336314}
\urldef\tempurl%
\url{https://doi.org/10.1145/2809695.2809711}
\showDOI{\tempurl}


\bibitem[{Chen, Yu-Hsin and Krishna, Tushar and Emer, Joel and Sze, Vivienne}(2016)]%
        {eyeriss}
\bibfield{author}{\bibinfo{person}{{Chen, Yu-Hsin and Krishna, Tushar and Emer, Joel and Sze, Vivienne}}.} \bibinfo{year}{{2016}}\natexlab{}.
\newblock \showarticletitle{{Eyeriss: An Energy-Efficient Reconfigurable Accelerator for Deep Convolutional Neural Networks}}. In \bibinfo{booktitle}{\emph{{IEEE International Solid-State Circuits Conference, ISSCC 2016, Digest of Technical Papers}}}. \bibinfo{pages}{{262--263}}.
\newblock


\bibitem[Cuervo et~al\mbox{.}(2010)]%
        {cuervo2010maui}
\bibfield{author}{\bibinfo{person}{Eduardo Cuervo}, \bibinfo{person}{Aruna Balasubramanian}, \bibinfo{person}{Dae-ki Cho}, \bibinfo{person}{Alec Wolman}, \bibinfo{person}{Stefan Saroiu}, \bibinfo{person}{Ranveer Chandra}, {and} \bibinfo{person}{Paramvir Bahl}.} \bibinfo{year}{2010}\natexlab{}.
\newblock \showarticletitle{MAUI: making smartphones last longer with code offload}. In \bibinfo{booktitle}{\emph{Proceedings of the 8th international conference on Mobile systems, applications, and services}}. ACM, \bibinfo{pages}{49--62}.
\newblock


\bibitem[Das et~al\mbox{.}(2018)]%
        {resource-aware-session-types}
\bibfield{author}{\bibinfo{person}{Ankush Das}, \bibinfo{person}{Jan Hoffmann}, {and} \bibinfo{person}{Frank Pfenning}.} \bibinfo{year}{2018}\natexlab{}.
\newblock \showarticletitle{Work Analysis with Resource-Aware Session Types}. In \bibinfo{booktitle}{\emph{Proceedings of the 33rd Annual ACM/IEEE Symposium on Logic in Computer Science}} (Oxford, United Kingdom) \emph{(\bibinfo{series}{LICS '18})}. \bibinfo{publisher}{ACM}, \bibinfo{address}{New York, NY, USA}, \bibinfo{pages}{305--314}.
\newblock
\showISBNx{978-1-4503-5583-4}
\urldef\tempurl%
\url{https://doi.org/10.1145/3209108.3209146}
\showDOI{\tempurl}


\bibitem[Di~Nucci et~al\mbox{.}(2017)]%
        {petra}
\bibfield{author}{\bibinfo{person}{Dario Di~Nucci}, \bibinfo{person}{Fabio Palomba}, \bibinfo{person}{Antonio Prota}, \bibinfo{person}{Annibale Panichella}, \bibinfo{person}{Andy Zaidman}, {and} \bibinfo{person}{Andrea De~Lucia}.} \bibinfo{year}{2017}\natexlab{}.
\newblock \showarticletitle{Software-based energy profiling of Android apps: Simple, efficient and reliable?}. In \bibinfo{booktitle}{\emph{2017 IEEE 24th International Conference on Software Analysis, Evolution and Reengineering (SANER)}}. \bibinfo{pages}{103--114}.
\newblock
\urldef\tempurl%
\url{https://doi.org/10.1109/SANER.2017.7884613}
\showDOI{\tempurl}


\bibitem[Dong and Zhong(2011)]%
        {Sesame}
\bibfield{author}{\bibinfo{person}{Mian Dong} {and} \bibinfo{person}{Lin Zhong}.} \bibinfo{year}{2011}\natexlab{}.
\newblock \showarticletitle{Self-constructive High-rate System Energy Modeling for Battery-powered Mobile Systems}. In \bibinfo{booktitle}{\emph{Proceedings of the 9th International Conference on Mobile Systems, Applications, and Services}} (Bethesda, Maryland, USA) \emph{(\bibinfo{series}{MobiSys '11})}. \bibinfo{publisher}{ACM}, \bibinfo{address}{New York, NY, USA}, \bibinfo{pages}{335--348}.
\newblock
\showISBNx{978-1-4503-0643-0}
\urldef\tempurl%
\url{https://doi.org/10.1145/1999995.2000027}
\showDOI{\tempurl}


\bibitem[Drucker et~al\mbox{.}(1997)]%
        {svr}
\bibfield{author}{\bibinfo{person}{Harris Drucker}, \bibinfo{person}{Christopher J.~C. Burges}, \bibinfo{person}{Linda Kaufman}, \bibinfo{person}{Alex~J. Smola}, {and} \bibinfo{person}{Vladimir Vapnik}.} \bibinfo{year}{1997}\natexlab{}.
\newblock \showarticletitle{Support Vector Regression Machines}.
\newblock In \bibinfo{booktitle}{\emph{Advances in Neural Information Processing Systems 9}}, \bibfield{editor}{\bibinfo{person}{M.~C. Mozer}, \bibinfo{person}{M.~I. Jordan}, {and} \bibinfo{person}{T.~Petsche}} (Eds.). \bibinfo{publisher}{MIT Press}, \bibinfo{pages}{155--161}.
\newblock
\urldef\tempurl%
\url{http://papers.nips.cc/paper/1238-support-vector-regression-machines.pdf}
\showURL{%
\tempurl}


\bibitem[Duan et~al\mbox{.}(2015)]%
        {automated-code-instrumentation}
\bibfield{author}{\bibinfo{person}{Yue Duan}, \bibinfo{person}{Mu Zhang}, \bibinfo{person}{Heng Yin}, {and} \bibinfo{person}{Yuzhe Tang}.} \bibinfo{year}{2015}\natexlab{}.
\newblock \showarticletitle{Privacy-Preserving Offloading of Mobile App to the Public Cloud}. In \bibinfo{booktitle}{\emph{7th {USENIX} Workshop on Hot Topics in Cloud Computing (HotCloud 15)}}. \bibinfo{publisher}{{USENIX} Association}, \bibinfo{address}{Santa Clara, CA}.
\newblock
\urldef\tempurl%
\url{https://www.usenix.org/conference/hotcloud15/workshop-program/presentation/duan}
\showURL{%
\tempurl}


\bibitem[Elbouchikhi(2018)]%
        {mlkit}
\bibfield{author}{\bibinfo{person}{Brahim Elbouchikhi}.} \bibinfo{year}{2018}\natexlab{}.
\newblock \bibinfo{title}{Introducing ML Kit}.
\newblock \bibinfo{howpublished}{\url{https://developers.googleblog.com/2018/05/introducing-ml-kit.html}}.
\newblock


\bibitem[Flinn and Satyanarayanan(1999)]%
        {PowerScope}
\bibfield{author}{\bibinfo{person}{Jason Flinn} {and} \bibinfo{person}{M. Satyanarayanan}.} \bibinfo{year}{1999}\natexlab{}.
\newblock \showarticletitle{PowerScope: A Tool for Profiling the Energy Usage of Mobile Applications}. In \bibinfo{booktitle}{\emph{Proceedings of the Second IEEE Workshop on Mobile Computer Systems and Applications}} \emph{(\bibinfo{series}{WMCSA '99})}. \bibinfo{publisher}{IEEE Computer Society}, \bibinfo{address}{Washington, DC, USA}, \bibinfo{pages}{2--}.
\newblock
\showISBNx{0-7695-0025-0}
\urldef\tempurl%
\url{http://dl.acm.org/citation.cfm?id=520551.837522}
\showURL{%
\tempurl}


\bibitem[{François Chollet and Others}(2015)]%
        {keras}
\bibfield{author}{\bibinfo{person}{{François Chollet and Others}}.} \bibinfo{year}{2015}\natexlab{}.
\newblock \bibinfo{title}{{Keras}}.
\newblock \bibinfo{howpublished}{\url{https://github.com/fchollet/keras}}.
\newblock


\bibitem[Friedman(2001)]%
        {gbr}
\bibfield{author}{\bibinfo{person}{Jerome~H. Friedman}.} \bibinfo{year}{2001}\natexlab{}.
\newblock \showarticletitle{Greedy Function Approximation: A Gradient Boosting Machine}.
\newblock \bibinfo{journal}{\emph{The Annals of Statistics}} \bibinfo{volume}{29}, \bibinfo{number}{5} (\bibinfo{year}{2001}), \bibinfo{pages}{1189--1232}.
\newblock
\showISSN{00905364}
\urldef\tempurl%
\url{http://www.jstor.org/stable/2699986}
\showURL{%
\tempurl}


\bibitem[Google(2023)]%
        {mlkitface}
\bibfield{author}{\bibinfo{person}{Google}.} \bibinfo{year}{2023}\natexlab{}.
\newblock \bibinfo{title}{Face Detection}.
\newblock \bibinfo{howpublished}{\url{https://developers.google.com/ml-kit/vision/face-detection}}.
\newblock


\bibitem[Guo et~al\mbox{.}(2018)]%
        {foggycache}
\bibfield{author}{\bibinfo{person}{Peizhen Guo}, \bibinfo{person}{Bo Hu}, \bibinfo{person}{Rui Li}, {and} \bibinfo{person}{Wenjun Hu}.} \bibinfo{year}{2018}\natexlab{}.
\newblock \showarticletitle{FoggyCache: Cross-Device Approximate Computation Reuse}. In \bibinfo{booktitle}{\emph{Proceedings of the 24th Annual International Conference on Mobile Computing and Networking}} (New Delhi, India) \emph{(\bibinfo{series}{MobiCom '18})}. \bibinfo{publisher}{ACM}, \bibinfo{address}{New York, NY, USA}, \bibinfo{pages}{19--34}.
\newblock
\showISBNx{978-1-4503-5903-0}
\urldef\tempurl%
\url{https://doi.org/10.1145/3241539.3241557}
\showDOI{\tempurl}


\bibitem[Guo and Hu(2018)]%
        {potluck}
\bibfield{author}{\bibinfo{person}{Peizhen Guo} {and} \bibinfo{person}{Wenjun Hu}.} \bibinfo{year}{2018}\natexlab{}.
\newblock \showarticletitle{Potluck: Cross-Application Approximate Deduplication for Computation-Intensive Mobile Applications}.
\newblock \bibinfo{journal}{\emph{ACM SIGPLAN Notices}}  \bibinfo{volume}{53} (\bibinfo{date}{03} \bibinfo{year}{2018}), \bibinfo{pages}{271--284}.
\newblock
\showISBNx{978-1-4503-4911-6}
\urldef\tempurl%
\url{https://doi.org/10.1145/3296957.3173185}
\showDOI{\tempurl}


\bibitem[Hoque et~al\mbox{.}(2015)]%
        {survey-energy-profiling}
\bibfield{author}{\bibinfo{person}{Mohammad~Ashraful Hoque}, \bibinfo{person}{Matti Siekkinen}, \bibinfo{person}{Kashif~Nizam Khan}, \bibinfo{person}{Yu Xiao}, {and} \bibinfo{person}{Sasu Tarkoma}.} \bibinfo{year}{2015}\natexlab{}.
\newblock \showarticletitle{Modeling, Profiling, and Debugging the Energy Consumption of Mobile Devices}.
\newblock \bibinfo{journal}{\emph{ACM Comput. Surv.}} \bibinfo{volume}{48}, \bibinfo{number}{3}, Article \bibinfo{articleno}{39} (\bibinfo{date}{Dec.} \bibinfo{year}{2015}), \bibinfo{numpages}{40}~pages.
\newblock
\showISSN{0360-0300}
\urldef\tempurl%
\url{https://doi.org/10.1145/2840723}
\showDOI{\tempurl}


\bibitem[Imai(2012)]%
        {imais-taskoffloading-android-salsa}
\bibfield{author}{\bibinfo{person}{Shigeru Imai}.} \bibinfo{year}{2012}\natexlab{}.
\newblock \emph{\bibinfo{title}{Task Offloading between Smartphones and Distributed Computational Resources}}.
\newblock \bibinfo{thesistype}{Master's\ thesis}. \bibinfo{school}{Rensselaer Polytechnic Institute}.
\newblock


\bibitem[Jiang et~al\mbox{.}(2018)]%
        {chameleon-video-analytics}
\bibfield{author}{\bibinfo{person}{Junchen Jiang}, \bibinfo{person}{Ganesh Ananthanarayanan}, \bibinfo{person}{Peter Bodík}, \bibinfo{person}{Siddhartha Sen}, {and} \bibinfo{person}{Ion Stoica}.} \bibinfo{year}{2018}\natexlab{}.
\newblock \showarticletitle{Chameleon: Scalable Adaptation of Video Analytics}. In \bibinfo{booktitle}{\emph{ACM Special Interest Group on Data Communication (SIGCOMM)}}.
\newblock
\urldef\tempurl%
\url{https://www.microsoft.com/en-us/research/publication/chameleon-video-analytics-scale-via-adaptive-configurations-cross-camera-correlations/}
\showURL{%
\tempurl}


\bibitem[Jung et~al\mbox{.}(2012)]%
        {DevScope}
\bibfield{author}{\bibinfo{person}{Wonwoo Jung}, \bibinfo{person}{Chulkoo Kang}, \bibinfo{person}{Chanmin Yoon}, \bibinfo{person}{Donwon Kim}, {and} \bibinfo{person}{Hojung Cha}.} \bibinfo{year}{2012}\natexlab{}.
\newblock \showarticletitle{DevScope: A Nonintrusive and Online Power Analysis Tool for Smartphone Hardware Components}. In \bibinfo{booktitle}{\emph{Proceedings of the Eighth IEEE/ACM/IFIP International Conference on Hardware/Software Codesign and System Synthesis}} (Tampere, Finland) \emph{(\bibinfo{series}{CODES+ISSS '12})}. \bibinfo{publisher}{ACM}, \bibinfo{address}{New York, NY, USA}, \bibinfo{pages}{353--362}.
\newblock
\showISBNx{978-1-4503-1426-8}
\urldef\tempurl%
\url{https://doi.org/10.1145/2380445.2380502}
\showDOI{\tempurl}


\bibitem[Kim et~al\mbox{.}(2015)]%
        {GPUPowerModel}
\bibfield{author}{\bibinfo{person}{Young~Geun Kim}, \bibinfo{person}{Minyong Kim}, \bibinfo{person}{Jae~Min Kim}, \bibinfo{person}{Minyoung Sung}, {and} \bibinfo{person}{Sung~Woo Chung}.} \bibinfo{year}{2015}\natexlab{}.
\newblock \showarticletitle{A Novel GPU Power Model for Accurate Smartphone Power Breakdown}.
\newblock \bibinfo{journal}{\emph{ETRI Journal}} \bibinfo{volume}{37}, \bibinfo{number}{1} (\bibinfo{year}{2015}), \bibinfo{pages}{157--164}.
\newblock
\urldef\tempurl%
\url{https://doi.org/10.4218/etrij.14.0113.1411}
\showDOI{\tempurl}
\showeprint{https://onlinelibrary.wiley.com/doi/pdf/10.4218/etrij.14.0113.1411}


\bibitem[{Kuhn}(2015)]%
        {gridsearch}
\bibfield{author}{\bibinfo{person}{M. {Kuhn}}.} \bibinfo{year}{2015}\natexlab{}.
\newblock \bibinfo{title}{{caret: Classification and Regression Training}}.
\newblock \bibinfo{howpublished}{Astrophysics Source Code Library}.
\newblock
\showeprint[ascl]{1505.003}


\bibitem[Lee et~al\mbox{.}(2014)]%
        {LeeModelGen}
\bibfield{author}{\bibinfo{person}{J. Lee}, \bibinfo{person}{H. Joe}, {and} \bibinfo{person}{H. Kim}.} \bibinfo{year}{2014}\natexlab{}.
\newblock \showarticletitle{Automated power model generation method for smartphones}.
\newblock \bibinfo{journal}{\emph{IEEE Transactions on Consumer Electronics}} \bibinfo{volume}{60}, \bibinfo{number}{2} (\bibinfo{date}{May} \bibinfo{year}{2014}), \bibinfo{pages}{190--197}.
\newblock
\showISSN{0098-3063}
\urldef\tempurl%
\url{https://doi.org/10.1109/TCE.2014.6851993}
\showDOI{\tempurl}


\bibitem[Li et~al\mbox{.}(2012)]%
        {cooks-distance}
\bibfield{author}{\bibinfo{person}{Jian Li}, \bibinfo{person}{Tomonori Nagayama}, \bibinfo{person}{Kirill~A. Mechitov}, {and} \bibinfo{person}{Billie F.~Spencer Jr.}} \bibinfo{year}{2012}\natexlab{}.
\newblock \showarticletitle{{Efficient campaign-type structural health monitoring using wireless smart sensors}}. In \bibinfo{booktitle}{\emph{Sensors and Smart Structures Technologies for Civil, Mechanical, and Aerospace Systems 2012}}, \bibfield{editor}{\bibinfo{person}{Masayoshi Tomizuka}, \bibinfo{person}{Chung-Bang Yun}, {and} \bibinfo{person}{Jerome~P. Lynch}} (Eds.), Vol.~\bibinfo{volume}{8345}. International Society for Optics and Photonics, \bibinfo{publisher}{SPIE}, \bibinfo{pages}{270 -- 280}.
\newblock
\urldef\tempurl%
\url{https://doi.org/10.1117/12.914860}
\showDOI{\tempurl}


\bibitem[{Lienhart} and {Maydt}(2002)]%
        {haar}
\bibfield{author}{\bibinfo{person}{R. {Lienhart}} {and} \bibinfo{person}{J. {Maydt}}.} \bibinfo{year}{2002}\natexlab{}.
\newblock \showarticletitle{An extended set of Haar-like features for rapid object detection}. In \bibinfo{booktitle}{\emph{Proceedings. International Conference on Image Processing}}, Vol.~\bibinfo{volume}{1}. \bibinfo{pages}{I--I}.
\newblock
\showISSN{1522-4880}
\urldef\tempurl%
\url{https://doi.org/10.1109/ICIP.2002.1038171}
\showDOI{\tempurl}


\bibitem[Maindonald(2006)]%
        {GAM}
\bibfield{author}{\bibinfo{person}{John Maindonald}.} \bibinfo{year}{2006}\natexlab{}.
\newblock \showarticletitle{Generalized Additive Models: An Introduction with R}.
\newblock \bibinfo{journal}{\emph{Journal of Statistical Software, Book Reviews}} \bibinfo{volume}{16}, \bibinfo{number}{3} (\bibinfo{year}{2006}), \bibinfo{pages}{1--2}.
\newblock
\showISSN{1548-7660}
\urldef\tempurl%
\url{https://doi.org/10.18637/jss.v016.b03}
\showDOI{\tempurl}


\bibitem[McCullough et~al\mbox{.}(2011)]%
        {HardwareModelsDoNotWork}
\bibfield{author}{\bibinfo{person}{John~C. McCullough}, \bibinfo{person}{Yuvraj Agarwal}, \bibinfo{person}{Jaideep Chandrashekar}, \bibinfo{person}{Sathyanarayan Kuppuswamy}, \bibinfo{person}{Alex~C. Snoeren}, {and} \bibinfo{person}{Rajesh~K. Gupta}.} \bibinfo{year}{2011}\natexlab{}.
\newblock \showarticletitle{Evaluating the Effectiveness of Model-based Power Characterization}. In \bibinfo{booktitle}{\emph{Proceedings of the 2011 USENIX Conference on USENIX Annual Technical Conference}} (Portland, OR) \emph{(\bibinfo{series}{USENIXATC'11})}. \bibinfo{publisher}{USENIX Association}, \bibinfo{address}{Berkeley, CA, USA}, \bibinfo{pages}{12--12}.
\newblock
\urldef\tempurl%
\url{http://dl.acm.org/citation.cfm?id=2002181.2002193}
\showURL{%
\tempurl}


\bibitem[Min et~al\mbox{.}(2015)]%
        {SmartwatchUserBattery}
\bibfield{author}{\bibinfo{person}{Chulhong Min}, \bibinfo{person}{Seungwoo Kang}, \bibinfo{person}{Chungkuk Yoo}, \bibinfo{person}{Jeehoon Cha}, \bibinfo{person}{Sangwon Choi}, \bibinfo{person}{Younghan Oh}, {and} \bibinfo{person}{Junehwa Song}.} \bibinfo{year}{2015}\natexlab{}.
\newblock \showarticletitle{Exploring Current Practices for Battery Use and Management of Smartwatches}. In \bibinfo{booktitle}{\emph{Proceedings of the 2015 ACM International Symposium on Wearable Computers}} (Osaka, Japan) \emph{(\bibinfo{series}{ISWC '15})}. \bibinfo{publisher}{ACM}, \bibinfo{address}{New York, NY, USA}, \bibinfo{pages}{11--18}.
\newblock
\showISBNx{978-1-4503-3578-2}
\urldef\tempurl%
\url{https://doi.org/10.1145/2802083.2802085}
\showDOI{\tempurl}


\bibitem[Pathak et~al\mbox{.}(2011)]%
        {eProfSystemCall}
\bibfield{author}{\bibinfo{person}{Abhinav Pathak}, \bibinfo{person}{Y.~Charlie Hu}, \bibinfo{person}{Ming Zhang}, \bibinfo{person}{Paramvir Bahl}, {and} \bibinfo{person}{Yi-Min Wang}.} \bibinfo{year}{2011}\natexlab{}.
\newblock \showarticletitle{Fine-grained Power Modeling for Smartphones Using System Call Tracing}. In \bibinfo{booktitle}{\emph{Proceedings of the Sixth Conference on Computer Systems}} (Salzburg, Austria) \emph{(\bibinfo{series}{EuroSys '11})}. \bibinfo{publisher}{ACM}, \bibinfo{address}{New York, NY, USA}, \bibinfo{pages}{153--168}.
\newblock
\showISBNx{978-1-4503-0634-8}
\urldef\tempurl%
\url{https://doi.org/10.1145/1966445.1966460}
\showDOI{\tempurl}


\bibitem[Pedregosa et~al\mbox{.}(2011)]%
        {scikit}
\bibfield{author}{\bibinfo{person}{Fabian Pedregosa}, \bibinfo{person}{Ga\"{e}l Varoquaux}, \bibinfo{person}{Alexandre Gramfort}, \bibinfo{person}{Vincent Michel}, \bibinfo{person}{Bertrand Thirion}, \bibinfo{person}{Olivier Grisel}, \bibinfo{person}{Mathieu Blondel}, \bibinfo{person}{Peter Prettenhofer}, \bibinfo{person}{Ron Weiss}, \bibinfo{person}{Vincent Dubourg}, \bibinfo{person}{Jake Vanderplas}, \bibinfo{person}{Alexandre Passos}, \bibinfo{person}{David Cournapeau}, \bibinfo{person}{Matthieu Brucher}, \bibinfo{person}{Matthieu Perrot}, {and} \bibinfo{person}{\'{E}douard Duchesnay}.} \bibinfo{year}{2011}\natexlab{}.
\newblock \showarticletitle{Scikit-learn: Machine Learning in Python}.
\newblock \bibinfo{journal}{\emph{J. Mach. Learn. Res.}}  \bibinfo{volume}{12} (\bibinfo{date}{Nov.} \bibinfo{year}{2011}), \bibinfo{pages}{2825--2830}.
\newblock
\showISSN{1532-4435}
\urldef\tempurl%
\url{http://dl.acm.org/citation.cfm?id=1953048.2078195}
\showURL{%
\tempurl}


\bibitem[Pfenning and Griffith(2015)]%
        {polarized-session-types}
\bibfield{author}{\bibinfo{person}{Frank Pfenning} {and} \bibinfo{person}{Dennis Griffith}.} \bibinfo{year}{2015}\natexlab{}.
\newblock \showarticletitle{Polarized Substructural Session Types}. In \bibinfo{booktitle}{\emph{Foundations of Software Science and Computation Structures}}, \bibfield{editor}{\bibinfo{person}{Andrew Pitts}} (Ed.). \bibinfo{publisher}{Springer Berlin Heidelberg}, \bibinfo{address}{Berlin, Heidelberg}, \bibinfo{pages}{3--22}.
\newblock
\showISBNx{978-3-662-46678-0}


\bibitem[Preuveneers et~al\mbox{.}(2020)]%
        {s20041176}
\bibfield{author}{\bibinfo{person}{Davy Preuveneers}, \bibinfo{person}{Ilias Tsingenopoulos}, {and} \bibinfo{person}{Wouter Joosen}.} \bibinfo{year}{2020}\natexlab{}.
\newblock \showarticletitle{Resource Usage and Performance Trade-offs for Machine Learning Models in Smart Environments}.
\newblock \bibinfo{journal}{\emph{Sensors}} \bibinfo{volume}{20}, \bibinfo{number}{4} (\bibinfo{year}{2020}).
\newblock
\showISSN{1424-8220}
\urldef\tempurl%
\url{https://doi.org/10.3390/s20041176}
\showDOI{\tempurl}


\bibitem[Rahmati et~al\mbox{.}(2007)]%
        {BatteryDissatisfaction}
\bibfield{author}{\bibinfo{person}{Ahmad Rahmati}, \bibinfo{person}{Angela Qian}, {and} \bibinfo{person}{Lin Zhong}.} \bibinfo{year}{2007}\natexlab{}.
\newblock \showarticletitle{Understanding Human-battery Interaction on Mobile Phones}. In \bibinfo{booktitle}{\emph{Proceedings of the 9th International Conference on Human Computer Interaction with Mobile Devices and Services}} (Singapore) \emph{(\bibinfo{series}{MobileHCI '07})}. \bibinfo{publisher}{ACM}, \bibinfo{address}{New York, NY, USA}, \bibinfo{pages}{265--272}.
\newblock
\showISBNx{978-1-59593-862-6}
\urldef\tempurl%
\url{https://doi.org/10.1145/1377999.1378017}
\showDOI{\tempurl}


\bibitem[{Rajaraman} et~al\mbox{.}(2014)]%
        {energy-mobile-video-streaming}
\bibfield{author}{\bibinfo{person}{S.~V. {Rajaraman}}, \bibinfo{person}{M. {Siekkinen}}, {and} \bibinfo{person}{M.~A. {Hoque}}.} \bibinfo{year}{2014}\natexlab{}.
\newblock \showarticletitle{Energy consumption anatomy of live video streaming from a smartphone}. In \bibinfo{booktitle}{\emph{2014 IEEE 25th Annual International Symposium on Personal, Indoor, and Mobile Radio Communication (PIMRC)}}. \bibinfo{pages}{2013--2017}.
\newblock
\urldef\tempurl%
\url{https://doi.org/10.1109/PIMRC.2014.7136502}
\showDOI{\tempurl}


\bibitem[Reif et~al\mbox{.}(2021)]%
        {tpu-power-model}
\bibfield{author}{\bibinfo{person}{Stefan Reif}, \bibinfo{person}{Benedict Herzog}, \bibinfo{person}{Judith Hemp}, \bibinfo{person}{Wolfgang Schr\"{o}der-Preikschat}, {and} \bibinfo{person}{Timo H\"{o}nig}.} \bibinfo{year}{2021}\natexlab{}.
\newblock \showarticletitle{AI Waste Prevention: Time and Power Estimation for Edge Tensor Processing Units: Poster}. In \bibinfo{booktitle}{\emph{Proceedings of the Twelfth ACM International Conference on Future Energy Systems}} (Virtual Event, Italy) \emph{(\bibinfo{series}{e-Energy '21})}. \bibinfo{publisher}{Association for Computing Machinery}, \bibinfo{address}{New York, NY, USA}, \bibinfo{pages}{300–301}.
\newblock
\showISBNx{9781450383332}
\urldef\tempurl%
\url{https://doi.org/10.1145/3447555.3466579}
\showDOI{\tempurl}


\bibitem[Romero et~al\mbox{.}(2021)]%
        {llama}
\bibfield{author}{\bibinfo{person}{Francisco Romero}, \bibinfo{person}{Mark Zhao}, \bibinfo{person}{Neeraja~J. Yadwadkar}, {and} \bibinfo{person}{Christos Kozyrakis}.} \bibinfo{year}{2021}\natexlab{}.
\newblock \showarticletitle{Llama: A Heterogeneous and Serverless Framework for Auto-Tuning Video Analytics Pipelines}. In \bibinfo{booktitle}{\emph{Proceedings of the ACM Symposium on Cloud Computing}} (Seattle, WA, USA) \emph{(\bibinfo{series}{SoCC '21})}. \bibinfo{publisher}{Association for Computing Machinery}, \bibinfo{address}{New York, NY, USA}, \bibinfo{pages}{1–17}.
\newblock
\showISBNx{9781450386388}
\urldef\tempurl%
\url{https://doi.org/10.1145/3472883.3486972}
\showDOI{\tempurl}


\bibitem[Sandur et~al\mbox{.}(2022)]%
        {jarvis}
\bibfield{author}{\bibinfo{person}{Atul Sandur}, \bibinfo{person}{ChanHo Park}, \bibinfo{person}{Stavros Volos}, \bibinfo{person}{Gul Agha}, {and} \bibinfo{person}{Myeongjae Jeon}.} \bibinfo{year}{2022}\natexlab{}.
\newblock \bibinfo{title}{Jarvis: Large-scale Server Monitoring with Adaptive Near-data Processing}.
\newblock
\newblock
\urldef\tempurl%
\url{https://doi.org/10.48550/ARXIV.2202.06021}
\showDOI{\tempurl}


\bibitem[Shankari et~al\mbox{.}(2018)]%
        {ZephyrUsingSoC}
\bibfield{author}{\bibinfo{person}{K. Shankari}, \bibinfo{person}{Jonathan Fürst}, \bibinfo{person}{Yawen Wang}, \bibinfo{person}{Philippe Bonnet}, \bibinfo{person}{David~E. Culler}, {and} \bibinfo{person}{Randy~H. Katz}.} \bibinfo{year}{2018}\natexlab{}.
\newblock \bibinfo{booktitle}{\emph{Zephyr: Simple, Ready-to-use Software-based Power Evaluation for Background Sensing Smartphone Applications}}.
\newblock \bibinfo{type}{{T}echnical {R}eport} UCB/EECS-2018-168. \bibinfo{institution}{EECS Department, University of California, Berkeley}.
\newblock
\urldef\tempurl%
\url{http://www2.eecs.berkeley.edu/Pubs/TechRpts/2018/EECS-2018-168.html}
\showURL{%
\tempurl}


\bibitem[Sharif et~al\mbox{.}(2021)]%
        {approxtuner}
\bibfield{author}{\bibinfo{person}{Hashim Sharif}, \bibinfo{person}{Yifan Zhao}, \bibinfo{person}{Maria Kotsifakou}, \bibinfo{person}{Akash Kothari}, \bibinfo{person}{Ben Schreiber}, \bibinfo{person}{E. Wang}, \bibinfo{person}{Yasmin Sarita}, \bibinfo{person}{Nathan Zhao}, \bibinfo{person}{Keyur Joshi}, \bibinfo{person}{V. Adve}, \bibinfo{person}{Sasa Misailovic}, {and} \bibinfo{person}{S. Adve}.} \bibinfo{year}{2021}\natexlab{}.
\newblock \showarticletitle{ApproxTuner: a compiler and runtime system for adaptive approximations}.
\newblock \bibinfo{journal}{\emph{Proceedings of the 26th ACM SIGPLAN Symposium on Principles and Practice of Parallel Programming}} (\bibinfo{year}{2021}).
\newblock


\bibitem[Shiftehfar et~al\mbox{.}(2015)]%
        {IMCM-full}
\bibfield{author}{\bibinfo{person}{Reza Shiftehfar}, \bibinfo{person}{Kirill Mechitov}, {and} \bibinfo{person}{Gul Agha}.} \bibinfo{year}{2015}\natexlab{}.
\newblock \showarticletitle{A Fine-Grained Adaptive Middleware Framework for Parallel Mobile Hybrid Cloud Applications}.
\newblock
\urldef\tempurl%
\url{https://doi.org/10.5176/2382-5669_ICT-BDCS15.08}
\showDOI{\tempurl}


\bibitem[Stamoulis et~al\mbox{.}(2018)]%
        {hyperpower}
\bibfield{author}{\bibinfo{person}{Dimitrios Stamoulis}, \bibinfo{person}{Ermao Cai}, \bibinfo{person}{Da-Cheng Juan}, {and} \bibinfo{person}{Diana Marculescu}.} \bibinfo{year}{2018}\natexlab{}.
\newblock \showarticletitle{HyperPower: Power- and memory-constrained hyper-parameter optimization for neural networks}. In \bibinfo{booktitle}{\emph{2018 Design, Automation Test in Europe Conference Exhibition (DATE)}}. \bibinfo{pages}{19--24}.
\newblock
\urldef\tempurl%
\url{https://doi.org/10.23919/DATE.2018.8341973}
\showDOI{\tempurl}


\bibitem[(student et~al\mbox{.}(2011)]%
        {demo:externalpowermonitor}
\bibfield{author}{\bibinfo{person}{Aaron~Schulman (student}, \bibinfo{person}{Thomas Schmid}, \bibinfo{person}{Prabal Dutta}, {and} \bibinfo{person}{Neil Spring)}.} \bibinfo{year}{2011}\natexlab{}.
\newblock \bibinfo{title}{Demo: Phone Power Monitoring with BattOr}.
\newblock
\newblock


\bibitem[Wang et~al\mbox{.}(2013)]%
        {BatteryTraces}
\bibfield{author}{\bibinfo{person}{C. Wang}, \bibinfo{person}{F. Yan}, \bibinfo{person}{Y. Guo}, {and} \bibinfo{person}{X. Chen}.} \bibinfo{year}{2013}\natexlab{}.
\newblock \showarticletitle{Power estimation for mobile applications with profile-driven battery traces}. In \bibinfo{booktitle}{\emph{International Symposium on Low Power Electronics and Design (ISLPED)}}. \bibinfo{pages}{120--125}.
\newblock
\urldef\tempurl%
\url{https://doi.org/10.1109/ISLPED.2013.6629277}
\showDOI{\tempurl}


\bibitem[Xu et~al\mbox{.}(2013)]%
        {Vedge}
\bibfield{author}{\bibinfo{person}{Fengyuan Xu}, \bibinfo{person}{Yunxin Liu}, \bibinfo{person}{Qun Li}, {and} \bibinfo{person}{Yongguang Zhang}.} \bibinfo{year}{2013}\natexlab{}.
\newblock \showarticletitle{V-edge: Fast Self-constructive Power Modeling of Smartphones Based on Battery Voltage Dynamics}. In \bibinfo{booktitle}{\emph{10th USENIX Symposium on Networked Systems Design and Implementation (NSDI 13)}}. \bibinfo{publisher}{USENIX Association}, \bibinfo{address}{Lombard, IL}, \bibinfo{pages}{43--55}.
\newblock
\showISBNx{978-1-931971-00-3}
\urldef\tempurl%
\url{https://www.usenix.org/conference/nsdi13/technical-sessions/presentation/xu_fengyuan}
\showURL{%
\tempurl}


\bibitem[Yang et~al\mbox{.}(2016)]%
        {FineABR}
\bibfield{author}{\bibinfo{person}{Jian Yang}, \bibinfo{person}{Yongyi Ran}, \bibinfo{person}{Shuangwu Chen}, \bibinfo{person}{Weiping Li}, {and} \bibinfo{person}{Lajos Hanzo}.} \bibinfo{year}{2016}\natexlab{}.
\newblock \showarticletitle{Online Source Rate Control for Adaptive Video Streaming Over HSPA and LTE-Style Variable Bit Rate Downlink Channels}.
\newblock \bibinfo{journal}{\emph{IEEE Transactions on Vehicular Technology}} \bibinfo{volume}{65}, \bibinfo{number}{2} (\bibinfo{year}{2016}), \bibinfo{pages}{643--657}.
\newblock
\urldef\tempurl%
\url{https://doi.org/10.1109/TVT.2015.2398515}
\showDOI{\tempurl}


\bibitem[Yang et~al\mbox{.}(2017)]%
        {eyeriss-memory-neuralnets}
\bibfield{author}{\bibinfo{person}{Tien-Ju Yang}, \bibinfo{person}{Yu-Hsin Chen}, \bibinfo{person}{Joel Emer}, {and} \bibinfo{person}{Vivienne Sze}.} \bibinfo{year}{2017}\natexlab{}.
\newblock \showarticletitle{A method to estimate the energy consumption of deep neural networks}. In \bibinfo{booktitle}{\emph{2017 51st Asilomar Conference on Signals, Systems, and Computers}}. \bibinfo{pages}{1916--1920}.
\newblock
\urldef\tempurl%
\url{https://doi.org/10.1109/ACSSC.2017.8335698}
\showDOI{\tempurl}


\bibitem[Yao et~al\mbox{.}(2020)]%
        {nn-offloading}
\bibfield{author}{\bibinfo{person}{Shuochao Yao}, \bibinfo{person}{Jinyang Li}, \bibinfo{person}{Dongxin Liu}, \bibinfo{person}{Tianshi Wang}, \bibinfo{person}{Shengzhong Liu}, \bibinfo{person}{Huajie Shao}, {and} \bibinfo{person}{T. Abdelzaher}.} \bibinfo{year}{2020}\natexlab{}.
\newblock \showarticletitle{Deep compressive offloading: speeding up neural network inference by trading edge computation for network latency}.
\newblock \bibinfo{journal}{\emph{Proceedings of the 18th Conference on Embedded Networked Sensor Systems}} (\bibinfo{year}{2020}).
\newblock


\bibitem[Yoon et~al\mbox{.}(2012)]%
        {AppScope}
\bibfield{author}{\bibinfo{person}{Chanmin Yoon}, \bibinfo{person}{Dongwon Kim}, \bibinfo{person}{Wonwoo Jung}, \bibinfo{person}{Chulkoo Kang}, {and} \bibinfo{person}{Hojung Cha}.} \bibinfo{year}{2012}\natexlab{}.
\newblock \showarticletitle{AppScope: Application Energy Metering Framework for Android Smartphone Using Kernel Activity Monitoring}. In \bibinfo{booktitle}{\emph{Presented as part of the 2012 {USENIX} Annual Technical Conference ({USENIX} {ATC} 12)}}. \bibinfo{publisher}{{USENIX}}, \bibinfo{address}{Boston, MA}, \bibinfo{pages}{387--400}.
\newblock
\showISBNx{978-931971-93-5}
\urldef\tempurl%
\url{https://www.usenix.org/conference/atc12/technical-sessions/presentation/yoon}
\showURL{%
\tempurl}


\bibitem[Zhang and Lee(2011)]%
        {batterySoC}
\bibfield{author}{\bibinfo{person}{Jingliang Zhang} {and} \bibinfo{person}{Jay Lee}.} \bibinfo{year}{2011}\natexlab{}.
\newblock \showarticletitle{A review on prognostics and health monitoring of Li-ion battery}.
\newblock \bibinfo{journal}{\emph{Journal of power sources}} \bibinfo{volume}{196}, \bibinfo{number}{15} (\bibinfo{year}{2011}), \bibinfo{pages}{6007--6014}.
\newblock


\bibitem[Zhang et~al\mbox{.}(2010)]%
        {PowerTutor}
\bibfield{author}{\bibinfo{person}{L. Zhang}, \bibinfo{person}{B. Tiwana}, \bibinfo{person}{R.~P. Dick}, \bibinfo{person}{Z. Qian}, \bibinfo{person}{Z.~M. Mao}, \bibinfo{person}{Z. Wang}, {and} \bibinfo{person}{L. Yang}.} \bibinfo{year}{2010}\natexlab{}.
\newblock \showarticletitle{Accurate online power estimation and automatic battery behavior based power model generation for smartphones}. In \bibinfo{booktitle}{\emph{2010 IEEE/ACM/IFIP International Conference on Hardware/Software Codesign and System Synthesis (CODES+ISSS)}}. \bibinfo{pages}{105--114}.
\newblock


\bibitem[Zhang et~al\mbox{.}(2015)]%
        {multicore-cpu-power-modeling}
\bibfield{author}{\bibinfo{person}{Yifan Zhang}, \bibinfo{person}{Yunxin Liu}, \bibinfo{person}{Li Zhuang}, \bibinfo{person}{Xuanzhe Liu}, \bibinfo{person}{Feng Zhao}, {and} \bibinfo{person}{Qun Li}.} \bibinfo{year}{2015}\natexlab{}.
\newblock \bibinfo{booktitle}{\emph{Accurate CPU Power Modeling for Multicore Smartphones}}.
\newblock \bibinfo{type}{{T}echnical {R}eport} MSR-TR-2015-9.
\newblock
\urldef\tempurl%
\url{https://www.microsoft.com/en-us/research/publication/accurate-cpu-power-modeling-for-multicore-smartphones/}
\showURL{%
\tempurl}


\bibitem[Zou and Hastie(2005)]%
        {lassolimit}
\bibfield{author}{\bibinfo{person}{Hui Zou} {and} \bibinfo{person}{Trevor Hastie}.} \bibinfo{year}{2005}\natexlab{}.
\newblock \showarticletitle{Regularization and variable selection via the elastic net (vol B 67, pg 301, 2005)}.
\newblock \bibinfo{journal}{\emph{Journal of the Royal Statistical Society Series B}}  \bibinfo{volume}{67} (\bibinfo{date}{02} \bibinfo{year}{2005}), \bibinfo{pages}{768--768}.
\newblock
\urldef\tempurl%
\url{https://doi.org/10.1111/j.1467-9868.2005.00527.x}
\showDOI{\tempurl}


\end{thebibliography}

\appendix

\begin{Appendix}
\section{Discussion}
\label{app:discuss}
\subsection{Scaling to Large Number of Mobile Devices}
The server can concurrently process logs from multiple mobile devices and \name{} clients. It can be easily scaled by provisioning more cloud servers to carry out the training jobs. This is because each of the collected training logs is specific to a mobile device, and thus needs to be trained independently of the other devices sending data to the server. There is no shared state or data communication between training jobs across clients, so each job can be mapped to a server node. The training jobs are not currently distributed as we execute each job entirely on a server node. In the future, we can consider distributing the training jobs for faster training. 
In case of retraining the prediction model for a client, we currently adopt a simple policy of considering each training job to be stateless, i.e., it does not accumulate the training data from previous runs for the client. This simplifies the need to train on the latest data which captures the hardware and application conditions on the device. We can extend the policy to incorporate more sophisticated strategies in the future. 

\subsection{Using \name{} During Inference}
\label{sec:escope-inference}
\name{} needs to provide \actor{}-level power predictions to a placement framework, which enables making energy-aware placement decisions. For this purpose, we leverage the notion of \emph{adaptation epochs} (time interval at which adaptation decisions are made). Once the trained \name{} model is deployed on the mobile device for inference, it is used to make \actor{}-level power predictions for the next adaptation epoch, based on the observed \actor{} execution time in the previous epoch. \name{} is queried by a placement framework to make placement decision per-\actor{} for the upcoming epoch, to decide whether to execute each \actor{} locally on the device or remote server.

If the input stream content changes very quickly, then the adaptation epoch size may need to be configured to a small enough duration so that the system can keep up with the input changes and not be stuck in a sub-optimal placement plan. However, doing so requires frequently running \name{} inference on the device, thereby increasing inference overhead. Thus, the effectiveness of \name{} is maintained as long as the epoch duration is long enough to avoid making power predictions and invoking placement decisions too frequently. We note that input changes occur slowly in our target scenarios, in the order of 10s of seconds~\cite{glimpse-continuous-face-track-data}. This allows our adaptation epoch sizes to be long enough (our evaluation was done for 1 second epoch size in Section~\ref{sec:evaluation} to stress the system but it can be configured for a longer period) so that the placement framework does not query for the \actor{} power costs too frequently, while still keeping up with the rate at which input stream changes occur. E.g., for face recognition using within a social media application such as Snapchat, we use a video stream where the mobile user is using the front camera to classify their facial expressions and the stream is steady as long as the user is in front of the camera. Similarly video conferencing applications where the background blurring occurs, has a steady input stream for the duration of the call which can be in the order of several minutes.

There is a key challenge with using the trained \name{} prediction model during inference. The input features for the trained model are the \actor{} active durations within a battery discharge interval. However, this means during inference, we need to wait for a discharge interval to occur (which can be $\sim$140 seconds for real-world applications we evaluated), so the predictions may not keep up with the change in input data, which can be in the order of 10s of seconds. To resolve this issue, we propose introducing an additional trained linear regressor which simply maps the \actor{} active duration within a 1 second segment, to the corresponding total \actor{} active duration within a battery discharge interval. E.g., if during training an \actor{} is determined to execute for 400 ms in each segment of 1 second duration, then the trained model outputs that the \actor{} was active for a total of 40\% of the discharge interval duration, if there exists a linear mapping from the \actor{} active duration within a segment and the corresponding duration within a discharge interval during training. In fact, analysis of our data collected for real-world applications suggests the correlation coefficient between \actor{} active duration within a segment and total \actor{} active duration within a discharge interval is between 0.82-0.93. Thus, as a first step towards enabling \name{} inference at fine-grained time scales, we propose using the simplified linear regression model for resolving the mismatch in input features for training vs. inference. For future work, we will analyze the data for more real-world applications and the use of more powerful prediction models beyond the linear regressor, to find more accurate mappings between segment-level and discharge interval-level \actor{} active durations.  

We can safely assume that input streams change slowly enough for our target scenarios, so that power predictions and associated placement adaptation decisions can keep up with the input. 

\subsection{Integration with a placement framework.}
\name{} is focused on addressing challenges for fine-grained power prediction for mobile video application operators. In our previous work Jarvis~\cite{jarvis}, we proposed a fine-grained operator-level placement mechanism for streaming analytics queries in datacenter server monitoring pipelines. This work can be applied to video analytics applications, since such applications can also be modeled as directed acyclic graph (DAG) of operators (similar to server monitoring queries) with vertices indicating the analytics operators and edges denoting dataflow dependencies between operators. The placement mechanism in Jarvis used a compute budget constraint as input and partitioned the application operators with the goal to minimize the cost of network data transfer from the servers in monitoring pipelines, subject to the available compute budget on the server node. We can extend such a framework to also incorporate an additional constraint in the optimization problem, which is the energy budget for mobile video analytics applications. Thus, the new placement solution can adapt the placement policy of video analytics operators to maximize the reduction in network transfer costs from mobile devices, without violating the compute and energy resource budget constraints. 

Note that we have so far focused on the power cost of using compute resources on the mobile device. However, maximizing compute resource utilization on the mobile device can incur high power costs, while minimizing compute resource use on the device can also incur high power costs depending on network conditions on the mobile device. E.g., using cellular connection to transport data to edge servers could be more energy-intensive compared to using WiFi connection~\cite{cuervo2010maui}. To address this issue, based on previous works which estimate energy costs based on the size of data transferred over the network~\cite{PowerTutor}, we rely on a critical assumption that reducing the data size for network transfer also reduces the network power costs. Combined with the property of analytics operators which result in intermediate data size reduction as we process more operators in the computational graph, we can leverage the placement algorithm in Jarvis, to identify a placement configuration which meets the compute and energy budget constraints. Integrating the concepts we discussed above, into a comprehensive, energy-aware placement solution for mobile devices, is an interesting direction for future work.


\end{Appendix}

\end{document}